\documentclass[a4paper,11pt]{article}
\usepackage{jheppub} 
\usepackage{lineno}
\usepackage{float}
\usepackage{jheppub}
\usepackage{lineno}
\nolinenumbers
\usepackage{booktabs}
\usepackage{hyperref}
\usepackage{longtable}
\usepackage{multirow}
\usepackage{caption}
\usepackage{slashed}
\usepackage{tikz}
\usepackage{tikz-feynman}
\tikzfeynmanset{compat=1.1.0} 
\usepackage{subcaption} 

\allowdisplaybreaks

\title{Displaced tracks as a signal of dark matter co-annihilation: a case study using inert doublet model at $e^+e^-$ collider.}

\author[1]{Hrishikesh Deka,}
\author[1]{Subhaditya Bhattacharya,}
\affiliation[1]{Department of Physics, Indian Institute of Technology, Guwahati, Assam-781039, India}
\author[2,3]{Amir Subba,}
\affiliation[2]{Wilczek Quantum Center, Shanghai Institute for Advanced Studies, Shanghai 201315, China}
\affiliation[3]{University of Science and Technology of China, Hefei 230026, China}
\author[4]{Ritesh K. Singh}
\affiliation[4]{Department of Physical Sciences, Indian Institute of Science Education and Research Kolkata, Mohanpur, West Bengal 741246, India}

\emailAdd{hrishikesh.deka@iitg.ac.in, subhab@iitg.ac.in, amirsubba@ustc.edu.cn, ritesh.singh@iiserkol.ac.in}

\abstract{Dark sector containing heavy (charged) particles with small mass splitting with dark matter (DM) can co-annihilate and produce correct relic density, evading direct detection bounds. In such compressed spectra, the charged state can be long-lived and decay to DM along with a very soft charged particle, giving rise to distinctive disappearing or kink track signatures at colliders. As an example, we consider the Inert Doublet Model (IDM), where the charged scalar $H^\pm$ can be nearly degenerate with the neutral DM state ($H^0$), and produce a soft pion ($\pi^\pm$) when it decays.  
We investigate this signature at a future $e^+e^-$ collider, focusing on electroweak $H^+H^-$ pair production and its displaced decays. By varying the mass splitting, we explore the transition from prompt soft-particle signatures to long-lived and disappearing-track signatures. Our analysis demonstrates the potential of future lepton colliders to probe the interplay between the charged-particle lifetime, its decay probability within the tracking volume, and the production rate sensitivity to the compressed DM scenarios over a broad range of masses, allowed by cosmological constraints. 
}

\begin{document}
\maketitle
\flushbottom

\section{Introduction}
\label{sec:intro}

The existence of dark matter (DM) is among the most compelling evidences for physics beyond the Standard Model (SM). A confluence of astrophysical and cosmological observations including galactic rotation curves~\cite{Sofue:2000jx}, gravitational lensing~\cite{Clowe:2006eq}, and precision measurements of the cosmic microwave background by WMAP and Planck~\cite{Planck:2018nkj} firmly establishes that approximately $27\%$ of the energy budget of our Universe is composed of a non-luminous, non-baryonic component. Unfortunately, the properties of DM can't be satisfied by any SM particle, and calls for physics beyond the SM. Despite this overwhelming indirect evidence, the fundamental nature of DM remains elusive: no direct  detection experiment has observed a DM-induced nuclear or electron recoil, no prompt collider signal excess or no indirect detection search  has unambiguously identified a DM annihilation signal.

Among the broad landscape of DM candidates, weakly interacting massive particles (WIMPs)  remain theoretically well-motivated. A particle with electroweak-scale mass and weak-strength interactions naturally reproduces the observed relic abundance through the thermal freeze-out mechanism. WIMPs arise naturally in a variety of SM extensions and predict observable signals at collider as well as direct/indirect detection experiments. Non observation of such a signal puts a bound on the DM mass and its couplings to the visible sector particles, eliminating a large parameter space of many of the simple WIMP scenarios.

Whether the dark sector particles share a SM charge is yet to be known; several possibilities in such directions have been chalked out. A particularly minimal and predictive realization of the WIMP paradigm is the Inert Doublet  Model (IDM), in which the SM is extended by a single 
$\mathcal{Z}_2$-odd $SU(2)_L$ scalar doublet ~\cite{Barbieri:2006dq, Cirelli:2005uq}. The discrete $\mathcal{Z}_2$ symmetry, under which all SM fields are even and the new doublet is odd, stabilizes the lightest component of the new doublet against decay, rendering it a viable DM candidate. Crucially, the choice of an $SU(2)_L$ doublet rather than a singlet is well-motivated on several grounds. First, a doublet representation automatically provides gauge interactions with the $W^\pm$ and $Z$ bosons, thereby generating the correct thermal relic density over a wide range of parameter space without the need for large Yukawa or Higgs-portal couplings that a singlet scenario would require. Second, and phenomenologically central to the present work, the doublet structure mandates the simultaneous presence of a neutral scalar $H^0$ (the DM candidate), a neutral pseudoscalar $A^0$, and a pair of charged scalars $H^\pm$. The near-degeneracy of these states, technically natural in the IDM and enforced by custodial symmetry considerations, gives rise to a small mass splitting $\Delta M \equiv M_{H^\pm} - M_{H^0}$, which plays a decisive role in the cosmological font, as well as in collider phenomenology. Whenever, the dark sector is charged under the SM, it is plagued with large annihilation, which can be tamed via kinematic effects with larger DM mass, together with DM co-annihilation with heavy dark sector particles. Effective co-annihilation requires a small mass splitting, which aids to the possibility of finding displaced vertex as we explore in this work.

The charged scalars $H^\pm$, once pair-produced, decay to the DM candidate $H^0$ and a soft charged pion or lepton via $H^\pm \to H^0 \pi^\pm$ or $H^\pm \to H^0 \ell^\pm \nu$. The pion interaction ($H^0H^\pm\pi^\pm$), is described within an effective framework following Ref.~\cite{PhysRevD.97.035011,PhysRevD.103.095006,Deka:2026gfo}.  When $\Delta M$ is small of order $\mathcal{O}(100~\text{MeV})$ to a few GeV, the visible decay products are too soft to satisfy conventional high-$p_T$ selection criteria, and the signal manifests as \emph{soft displaced tracks} accompanied by missing energy. Such soft tracks are low-momentum charged particle trajectories that, depending on 
$\Delta M$ and the corresponding decay length $c\tau$, may originate at a measurable displacement from the primary interaction vertex before the charged scalar decays. This topology is qualitatively distinct from the hard-object signatures exploited in standard mono-$X$ searches~\cite{CMS:2020bfa,CMS:2021edw,ATLAS:2022ihe,ATLAS:2024lpr,CMS:2021far} and provides sensitivity to a region of WIMP parameter space that is otherwise largely inaccessible.

The phenomenology of soft and disappearing tracks has been explored extensively at hadron colliders, most notably in the context of chargino production in supersymmetric models at the LHC~\cite{ATLAS:2022rme,ATLAS:2017oal,CMS:2018rea,CMS:2020atg}. However, the hadronic environment of the LHC imposes significant challenges from backgrounds, underlying event activity, and pileup, limiting sensitivity particularly at small $\Delta M$. Lepton colliders offer a considerably cleaner environment: the absence of QCD backgrounds, precise knowledge of the center-of-mass energy, and a well-defined initial state allow soft-track signatures to be exploited with much greater efficacy and with a lower kinematic threshold \cite{PhysRevLett.134.181802}.

In this work, we study the soft-track signature of the inert doublet model at a future $e^+e^-$ collider, in particular CLIC~\cite{CLICdp:2017vju} setting with $\sqrt{s}=3$ TeV. We focus on the pair production process $e^+e^- \to H^+H^-$, followed by the decay of each charged scalar to $H^0$ and a soft charged pion track in the regime of small mass splitting $\Delta M$. We demonstrate that the combination of soft-track reconstruction and missing-energy requirements provides a powerful handle on the IDM parameter space that is complementary to existing LHC bounds and direct detection constraints. We would also like to mention that the relation between $\Delta M$ and displaced vertex also indicates to a possible co-annihilation process through which DM gets saturated, offering an opportunity to correlate the collider and cosmological interpretation. Such interpretations can be made for any extended dark sector; which we have demonstrated here for the case of IDM.   

The remainder of the paper is organized as follows: In Section~\ref{sec:idm}, we discuss the IDM in detail along with current bounds on parameter space. In Section \ref{dm_analysis}, we present the dark matter phenomenology including relic density, direct detection. Section \ref{H_as_LLP} corresponds to the decay properties of the charged scalar and its long-lived signature . Next, the Section \ref{collider_analysis} is devoted to the collider study at 3 TeV CLIC, where we investigate the displaced-vertex, kink-track, and the disappearing-track signature arising from the long-lived $H^\pm$ states. Finally, we summarize our findings and conclude in Section \ref{conclusion}.

\section{The Model}
\label{sec:idm}
The Inert Doublet Model (IDM) extends the SM by adding a second scalar $SU(2)_L$ doublet $\Phi_2$ with hypercharge $Y=1/2$, which is odd under an unbroken $\mathcal{Z}_2$ symmetry, in addition to the $\mathcal{Z}_2$ even SM Higgs doublet ($\Phi_1$), also having $Y=1/2$ (following the convention $Q=T_3+Y$):
\[
\Phi_1 \rightarrow \Phi_1, \qquad \Phi_2 \rightarrow -\,\Phi_2 \,.
\]
 The doublet $\Phi_2$ does not acquire a vacuum expectation value (VEV) and does not couple to fermions and does not mix with the SM Higgs after the spontaneous electroweak symmetry breaking (EWSB), owing to the $\mathcal{Z}_2$. The two scalar doublets are given below~\cite{Branco:2011iw}:

\[
\Phi_1 =
\begin{pmatrix}
G^{+} \\
\frac{1}{\sqrt{2}}\,(v + h + i G^{0})
\end{pmatrix}, \qquad
\Phi_2 =
\begin{pmatrix}
H^{+} \\
\frac{1}{\sqrt{2}}\,(H^{0} + i A^{0})
\end{pmatrix}.
\]
Here, $G^0$ and $G^+$ represent the Goldstone bosons which are absorbed by the $Z$ and $W^+$ bosons to get their masses, respectively. $v$ is the standard VEV and $h$ is the physical Higgs Boson. In addition to the SM-like Higgs $h$, we have a pair of charged Higgs ($H^\pm$), a CP even neutral scal $H^0$ and a CP odd neutral scalar $A^0$.

The most general renormalizable and $\mathcal{Z}_2$-conserving scalar potential is
\begin{align}
V &= \mu_1^2 |\Phi_1|^2 + \mu_2^2 |\Phi_2|^2
+ \lambda_1 |\Phi_1|^4 + \lambda_2 |\Phi_2|^4
+ \lambda_3 |\Phi_1|^2|\Phi_2|^2 \nonumber \\
&\quad
+ \lambda_4 |\Phi_1^\dagger \Phi_2|^2
+ \frac{\lambda_5}{2} \left[(\Phi_1^\dagger \Phi_2)^2 +
(\Phi_1^\dagger \Phi_2)^{\!*2} \right] \,.
\end{align}

In above, $\mu_1^2<0, \lambda_1 >0$, whereas $\mu_2^2>0, \lambda_2 >0$. All parameters are real, ensuring CP conservation. The inert vacuum is defined by
\[
\langle \Phi_1 \rangle = \frac{v}{\sqrt{2}}, \qquad
\langle \Phi_2 \rangle = 0 .
\]
After the symmetry breaking, the scalar mass spectrum is given by
\begin{align}
M_h^2 &= 2\lambda_1 v^2 , ~~~~~~
M_{H^\pm}^2 = \mu_2^2 + \frac{1}{2}\lambda_3 v^2 ,  ~~~~~~
M_{H^0/A^0}^2 = \mu_2^2 + \frac{1}{2}(\lambda_3 + \lambda_4 \pm \lambda_5)v^2\,. 
\end{align}

It is convenient to define $\lambda_{L/S} = (\lambda_3 + \lambda_4 \pm \lambda_5)/2$. Throughout our analysis, we keep $\lambda_5<0$ just to make sure $M_{A^0}>M_{H^0}$ so that $H^0$ can be a potential DM candidate. However, for $\lambda_5>0$, the mass hierarchy is reversed, making $A^0$ the DM candidate.

\subsection*{Constraints on the parameters:}
There are several theoretical, experimental and observational constraints on the model parameters $\mu_{1,2}$ and $\lambda's$ or equivalently on the four masses, $\lambda_2$ and $\mu_2$. some important constraints are listed in the Table.\ref{tab:constraints}:

\begin{longtable}{|c|p{9.8cm}|}
\centering
\renewcommand{\arraystretch}{1.3}

\textbf{Constraint} & \textbf{Condition / Description} \\
\hline
\endfirsthead

\hline
\textbf{Constraint} & \textbf{Condition / Description} \\
\hline
\endfirsthead

\hline
\multicolumn{2}{r}{\textit{Continued on next page}} \\
\endfoot
\hline 
\caption{Summary of the theoretical and experimental constraints imposed on the IDM parameter space.}
\label{tab:constraints} \\
\endlastfoot
Vacuum Stability (BFB) &
$\lambda_{1,2}>0$,
\quad
$\lambda_3+\lambda_4-|\lambda_5|
>-2\sqrt{\lambda_1\lambda_2}$,
and inert vacuum condition:
$
\frac{\mu_2^2}{\sqrt{\lambda_2}}
\geq
\frac{\mu_1^2}{\sqrt{\lambda_1}}$ \cite{Ginzburg:2010wa,PhysRevD.18.2574,Kanemura:1993hm}

\\ & \\
\hline

Perturbativity &
All quartic couplings satisfy $|\lambda_i|\leq 4\pi$ to ensure validity of perturbation theory.
\\
\hline

Higgs Diphoton Signal Strength &
Charged scalar loop contribution to
$h\rightarrow\gamma\gamma$.
The measured signal strength is
$\mu_{\gamma\gamma}
=1.04^{+0.10}_{-0.09}$, 
which constrains $M_{H^\pm}$ and $\lambda_3$.
\cite{Posch:2010hx,ATLAS:2022tnm}
\\
\hline
Invisible Higgs Decay &
$\mathrm{BR}(h\rightarrow\mathrm{inv})<0.107$
\cite{ATLAS:2023tkt} at 95\% C.L. It is relevant when
$M_{H^0}<{M_h}/{2}.$
\\
\hline

Electroweak Precision Observables &
Oblique parameters constrained within 
\cite{PhysRevD.110.030001}: \\&
$S=-0.04\pm0.10,\qquad
T=0.01\pm0.12,\qquad
U=0.05\pm0.09.$
These are sensitive to the scalar mass splittings.
\\
\hline

LEP Scalar Searches &
Excluded region from LEP-II
\cite{OPAL:1999bfx,OPAL:2002lje}: \\&
$M_{H^0}<80~\mathrm{GeV},~~
M_{A^0}<100~\mathrm{GeV},~~
M_{H^\pm}>70~\mathrm{GeV}.$
\\
\hline

Gauge Boson Width Constraints &
To avoid additional decay channels of the $W$ and $Z$ bosons
\cite{L3:1999znj,CDF:1989qta,D0:2002wga,Novikov:1999af}: \\&
$M_{A^0}+M_{H^\pm}\geq M_W,~~
M_{H^0}+M_{H^\pm}\geq M_W,
$ \\&

$M_{H^0}+M_{A^0}\geq M_Z,~
2M_{H^\pm}\geq M_Z.$
\\
\end{longtable}

\section{Dark matter analysis}\label{dm_analysis}
Cosmological observations for example, the latest results from the Planck collaboration report that the observed relic density~\cite{Planck:2018vyg}  
\begin{equation}
    \Omega h^2=0.120\pm0.001\,,
\end{equation}
indicating about 26-27\% of the energy density of the universe is composed of DM. So, any model having a potential DM candidate must satisfy this relic density or otherwise may remain under-abundant or over abundant. Under abundance preferably leads to the assumption that there may exist more DM candidates. IDM has a potential DM candidate of WIMP type and the relic abundance is mainly governed by the annihilation and co-annihilation processes involving the dark sector scalars. The relic density is very sensitive to the mass splitting between the scalars. \par 
In the lower mass region, where ${H^0}$ (DM) mass is less than $W,Z$ or $h$ mass, $H^0$ mainly annihilates to $f \bar f$ pairs ($b\bar b$ is the dominant). As the $H^0$ mass crosses the $W, Z$ and $h$ mass threshold, the annihilation channels like $H^0H^0 \rightarrow WW, ~ZZ, ~hh$ become more efficient and strongly suppresses the relic density. 

In the higher mass region, $M_{H^0} \gg M_W$, the dominant annihilation channels are $H^0H^0\rightarrow WW,~ZZ,~hh$ which proceeds via both gauge and scalar mediated processes. However, in the limit of small $\lambda_L$ ($\mathcal{O}(10^{-2})$), the $H^0H^0\rightarrow hh$ channel contribution is very small. In this case, the dominant contributions come from the quartic gauge interaction involving two gauge bosons, along with the $t$- and $u$- channel exchange of $A^0$ (for $ZZ$) or $H^\pm$ (for $W^+W^-$). The annihilation cross section into $ZZ$ and $W^+W^-$ pair grows like \cite{LopezHonorez:2006gr}

\begin{equation} \label{zzww}
    \langle\sigma v\rangle_{ZZ}\sim \frac{\alpha_W^2(M_{A^0}-M_{H^0})^2}{M_Z^4} \sim \frac{\lambda_5^2}{M_{H^0}^2};
 ~~~~
     \langle\sigma v\rangle_{WW}\sim \frac{\alpha_W^2(M_{H^\pm}-M_{H^0})^2}{M_W^4} \sim \frac{(\lambda_4+\lambda_5)^2}{M_{H^0}^2}
\end{equation} 

Thus, these mass splittings play a crucial role not only in the co-annihilation as mentioned before, but also in the annihilation channels involving gauge bosons in the final state. Clearly, the large mass splitting correlates with a large quartic coupling region. Conversely, the annihilation cross section decreases in a nearly degenerate case, having a tussle with annihilation to gauge bosons.


\par In addition to the relic density, the observability of the dark matter  is subject to constraints from its experimental searches. These searches can be grouped into three main approaches: direct detection , indirect detection and collider searches.  Direct detection experiments are based on the scattering of dark matter with the nuclei of the detector material, and provide some of the most stringent bound on the dark matter \& SM interactions.  Some of the dedicated direct detection experiments are like \text{XENONnT}~\cite{XENON:2023cxc}, PandaX-4T~\cite{PandaX:2024qfu} , LUX-ZEPLIN (LZ)~\cite{LZ:2024zvo}, PICO~\cite{PICO:2019vsc} etc. In the indirect detection, it aims to probe the dark matter by searching the SM particles that are produced during the annihilation of dark matter, specifically at the centre of the galaxies. Experiments like Fermi Large Area Telescope (Fermi-LAT)~\cite{Hess:2021cdp}, H.E.S.S~\cite{Kerszberg:2023cup} have placed stringent constraint on the dark matter annihilation cross section. On the other hand, collider experiments like LHC provide a powerful environment to explore dark matter and beyond the SM physics. At the colliders, the dark matter particles can be produced in pairs or in association with the SM particles, leading to the missing transverse energy signature accompanied by SM objects.

In the context of the IDM, the dark matter candidate arises as the lightest neutral particle, stabilized by the discrete $\mathcal Z_2$ symmetry. Here, the $\mathcal Z_2$ odd sector particles are : the neutral CP-even scalar $H^0$, neutral CP-odd scalar $A^0$ and the charged Higgs $H^\pm$. Being charged, $H^\pm$ is automatically discarded, whereas the lightest among $H^0$ and $A^0$ serves as a potential dark matter candidate. For the dark matter phenomenology study, we scan the following input parameters in \texttt{micrOmegas}~\cite{Alguero:2023zol} in the following specified range:
\begin{align}\label{scanned_range}
 10~\text{GeV} \leq M_{H^0} < M_{A^0}, M_{H^+} \leq 1500~\text{GeV}, \qquad
0 < \lambda_2 < 4\pi, \qquad
|\lambda_{L}| \leq 0.5 .
\end{align}

\begin{figure}[H]
 \centering 
\includegraphics[width=.7\textwidth]{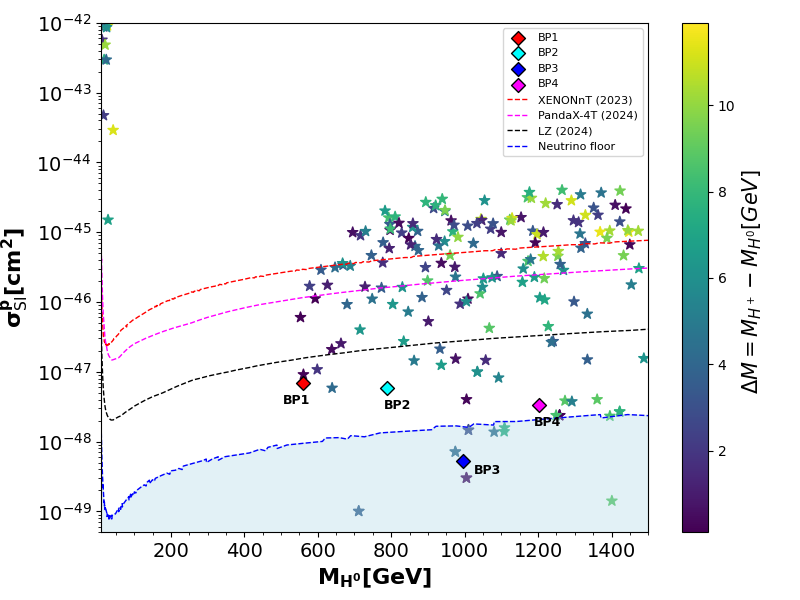}
 \caption{\label{xenon23}  Spin-independent cross-section as a function of DM mass ($M_{H_0}$). The '*' markers represent the points satisfying correct relic density. The selected benchmark points are shown with the diamond markers.}
\end{figure}

In Fig.~\ref{xenon23}, we show the parameter-space points that satisfy the observed correct relic density in the DM mass vs direct detection cross section ($\sigma_{\rm SI}$) plane. The red dashed line represents the XENONnT(2023) bound. Likewise, the PandaX-4T(2024) and LZ(2024) bounds are shown by the magenta and black dashed lines. Along with that the blue dashed line represents the statistical boundary where the neutrino backgrounds produce irreducible noise. The blue shaded region is what we call neutrino floor. In this region, it is extremely hard to distinguish the DM from the neutrino. The color bar represents the mass splitting between the $H^\pm$ and $H^0$ ($\Delta M =M_{H^\pm}-M_{H^0}$). Here, from the plot, one can see that we have a very small window for a viable WIMP like dark matter candidate. However, we choose four benchmark points viz. BP1, BP2, BP3 and BP4 satisfying all the theoretical constraints, satisfying relic and direct detection cross section (see Section \ref{collider_analysis} for details). Notice that we choose BP3 below the neutrino sensitivity bound. Although, in this region, direct detection experiments loose sensitivity, the collider production of inert scalars remains unsuppressed. So, the pair production followed by the decay may provide an important avenue to probe this at the collider, showing the complementarity between the cosmological search and the collider search. 

\begin{figure}[H]
 \centering 
\includegraphics[width=.49\textwidth]{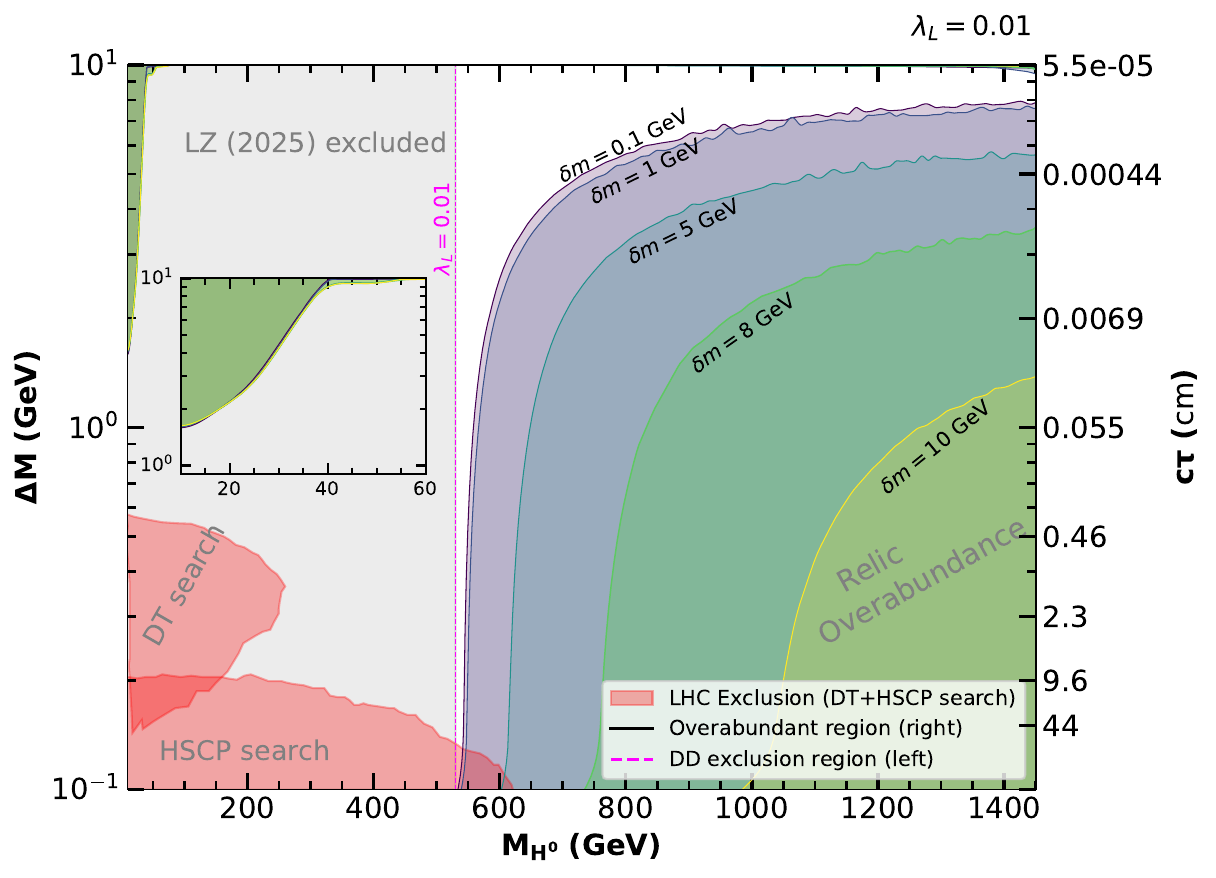}
\includegraphics[width=.49\textwidth]{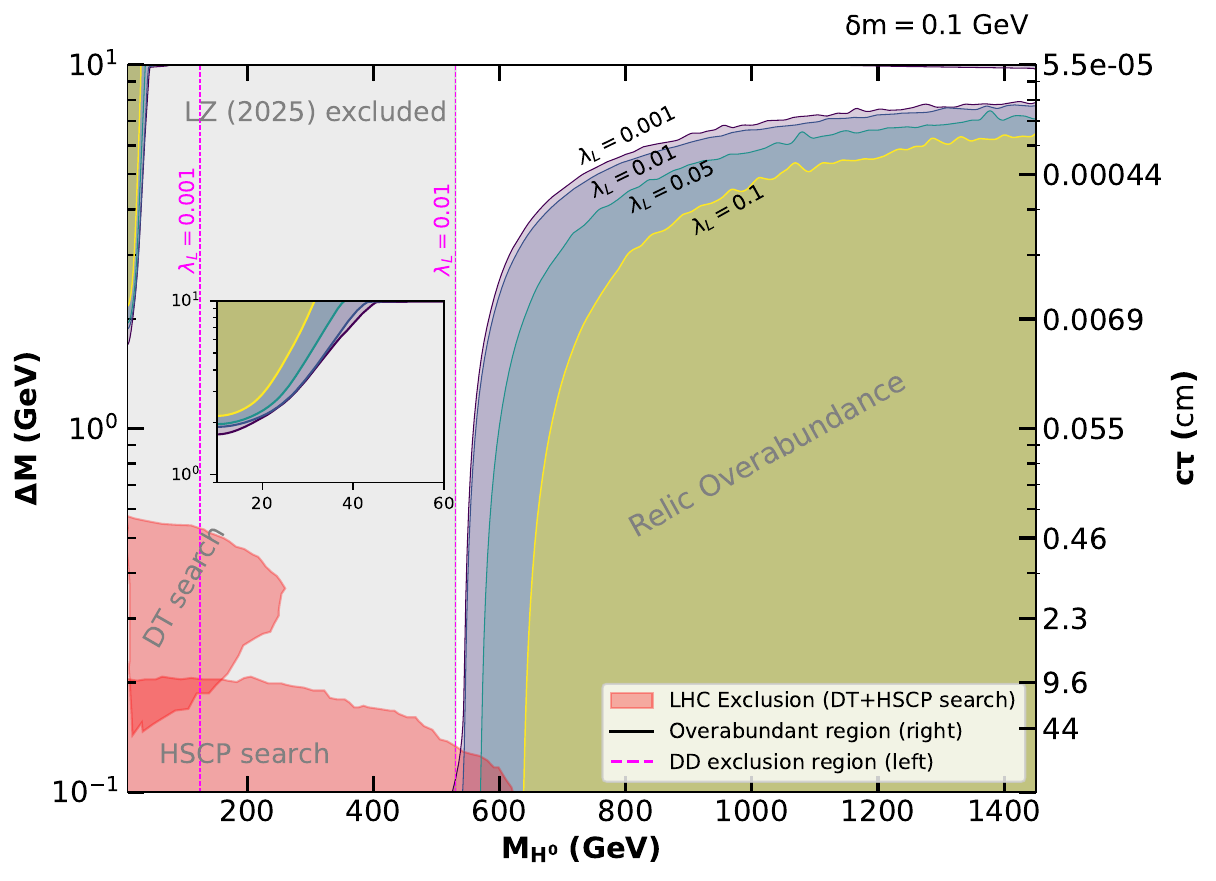}
 \caption{\label{delM}  Relic density and direct detection constraints in the $\Delta M-M_{H^0}$ plane for different benchmark choices of the scalar mass splitting $\delta m $ (left) and quartic coupling $\lambda_L$ (right). The shaded regions correspond to the relic overabundance, while the boundary serves as the correct relic allowed contour for each $\delta m$ (on the left) and each $\lambda_L$ (on the right). The
 gray shaded area, on the left of the magenta dashed lines on both plots indicates the exclusion limits from LZ (2024) bounds. The red shaded regions (in the left corner) are excluded by LHC disappearing track (DT) and Heavy Stable Charged Particle (HSCP) searches.}
\end{figure}

The dark matter relic density is highly sensitive to the mass splittings $\Delta M$ and $\delta m$, where we define $\delta m=M_{A^0}-M_{H^\pm}$. To investigate the impact of these splittings, we present the  $\Delta M$ vs $M_{H^0}$ parameter space in Fig.~\ref{delM}.  The left panel  shows the relic density and direct detection constraints in the $\Delta M$ vs $M_{H^0}$ plane. For this analysis, we scan the dark matter mass $M_{H^0}$ in the range as specified in the Eqn.~\eqref{scanned_range}. In addition, we choose different benchmark choices of the scalar splitting, namely $\delta m=0.1,~1,~5,~8$ and 10 GeV, which are considered independently to see their effects. The quartic coupling $\lambda_L$ is kept fixed at a  representative value of 0.01. The gray shaded region is excluded by the direct detection experiment LZ(2024). The right regions corresponding to each $\delta m$ contours represent the relic overabundance region, $\Omega h^2>0.12$. For small $\delta m$, the inert scalar states $A^0$ and $H^\pm$ becomes nearly degenerate, leading to a smaller annihilation rate into gauge bosons and consequently gives a larger relic abundance (see Eqn.~\eqref{zzww}). The co-annihilation effects are significant only in the low mass region, while they become suppressed for larger dark matter masses. This behavior is highlighted in the inset plot. Notably, in the the low mass regime, the relic density remains largely insensitive to $\delta m$, since the dark matter annihilation channels into gauge bosons are kinematically forbidden here.
Similarly, the right panel of Fig.~\ref{delM} shows the relic density and direct detection constraints for four representative $\lambda_L$ values 0.001, 0.01, 0.05, 0.1, while fixing $\delta m=0.1$ GeV. The region left to each vertical dashed magenta line represents the exclusion region by the direct detection experiments. Here, we show the exclusion boundaries only for $\lambda_L=0.001, ~0.01$, as for the other two $\lambda_L$  values, the whole region is excluded by the direct detection experiments. Like in the left panel , the regions to the right of each contour correspond to overabundance. In contrast to the dependence on $\delta m$, the relic density in the low mass region  exhibits a noticeable sensitivity to $\lambda_L$, since Higgs mediated annihilation channels are directly controlled by this coupling. In both plots, the red shaded region represents the current LHC exclusion region based on the disappearing track (DT) and the Heavy Stable Charged Particle (HSCP) search. Details of this exclusion analysis are discussed in the Section.~\ref{collider_analysis} 

\section{$H^\pm$ as long lived particle:}\label{H_as_LLP}
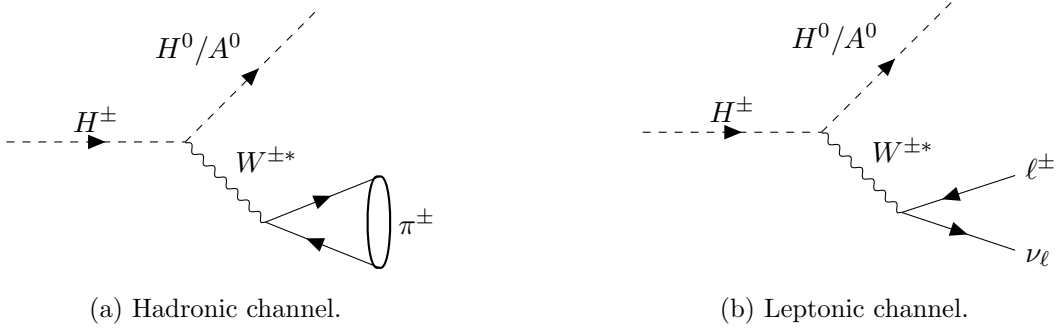
\begin{figure}[!t]
    \centering
    \begin{subfigure}{0.45\textwidth}
        \centering
        \begin{tikzpicture}
          \begin{feynman}
            \vertex (a) {};
            \vertex [right=2.5cm of a] (b);
            \vertex [above right=2.5cm of b] (c) {};
            \vertex [below right=1.5cm of b] (w);
            \vertex [right=1.5cm of w, yshift=0.6cm] (f1);
            \vertex [right=1.5cm of w, yshift=-0.6cm] (f2);

            \diagram* {
              (a) -- [charged scalar, edge label=\(H^\pm\)] (b),
              (b) -- [charged scalar, edge label=\(H^0/A^0\)] (c),
              (b) -- [boson, edge label=\(W^{\pm*}\)] (w),
              (w) -- [fermion] (f1),
              (w) -- [anti fermion] (f2),
            };

            \draw[thick] (f1) .. controls +(0.2,0.1) and +(0.2,-0.1) .. (f2)
                         .. controls +(-0.2,-0.1) and +(-0.2,0.1) .. (f1);
            \node at ($(f1)!0.5!(f2)+(.5,0)$) {\(\pi^\pm\)};
          \end{feynman}
        \end{tikzpicture}
        \caption{
        Hadronic channel.}
    \end{subfigure}%
    \hfill
    \begin{subfigure}{0.45\textwidth}
        \centering
        \begin{tikzpicture}
          \begin{feynman}
            \vertex (a) {};
            \vertex [right=2.5cm of a] (b);
            \vertex [above right=2.5cm of b] (c) {};
            \vertex [below right=1.5cm of b] (w);
            \vertex [right=1.5cm of w, yshift=0.6cm] (f1){$\ell^\pm$};
            \vertex [right=1.5cm of w, yshift=-0.6cm] (f2) {$\nu_\ell$};

            \diagram* {
              (a) -- [charged scalar, edge label=\(H^\pm\)] (b),
              (b) -- [charged scalar, edge label=\(H^0/A^0\)] (c),
              (b) -- [boson, edge label=\(W^{\pm*}\)] (w),
              (w) -- [anti fermion] (f1),
              (w) -- [fermion] (f2),
            };
          \end{feynman}
        \end{tikzpicture}
        \caption{
        Leptonic channel.}
    \end{subfigure}
        \caption{Feynman diagrams of $H^\pm$ decays with mass splitting between the charged and neutral scalars, $\Delta M <1$ GeV.}
    \label{fig:w-decays}
\end{figure}

\begin{figure}[H]
 \centering 
\includegraphics[width=.48\textwidth]{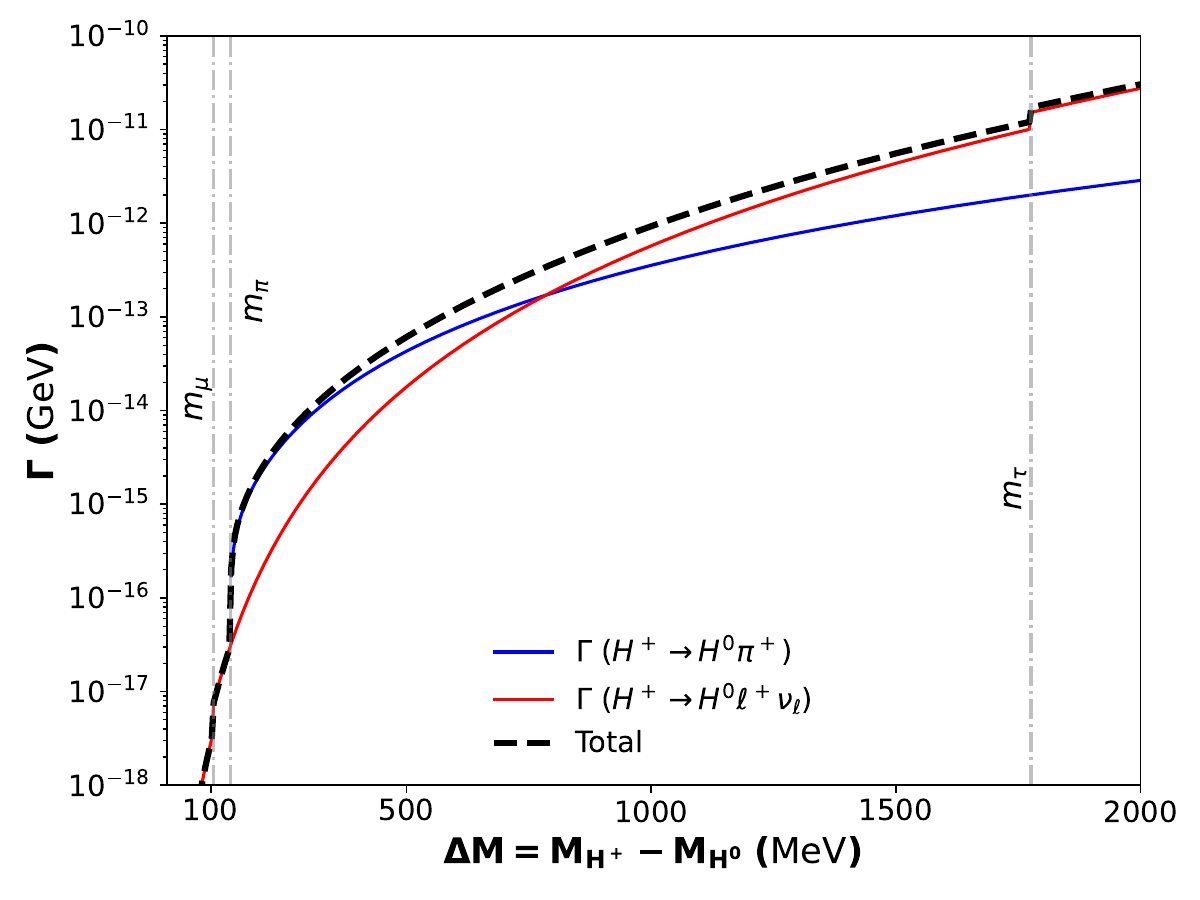}
\includegraphics[width=.48\textwidth]{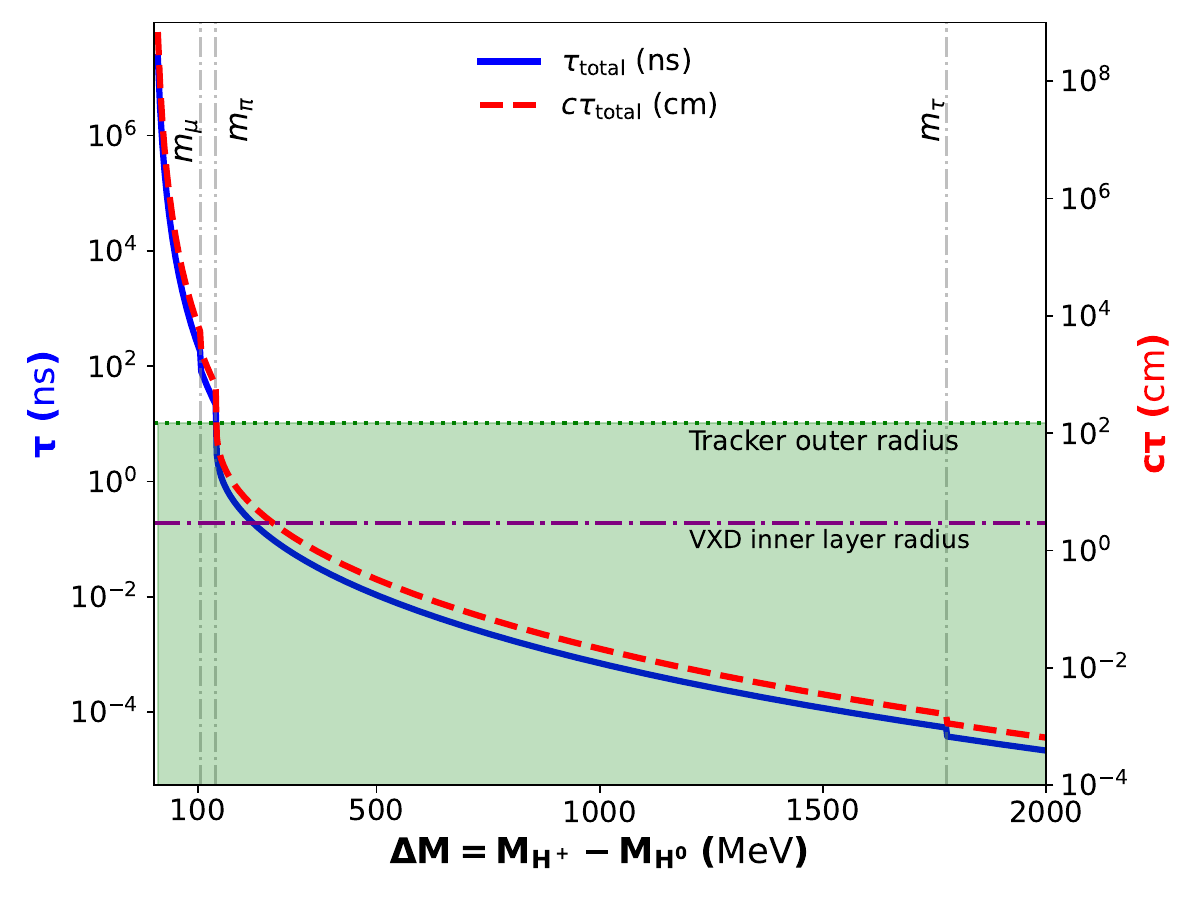}
 \caption{\label{decay_width1} \textit{Left:} Decay width of $H^+$ vs mass splitting $\Delta M$ for the hadronic (blue), leptonic (red) and hadronic+leptonic (black) . \textit{Right:} The blue (thick) and red (dashed) lines show the variation of lifetime $\tau$~(ns)  and  proper decay length $c\tau$~(cm) with mass splitting $\Delta M$ respectively.  }
\end{figure}

 We focus on the compressed spectrum between the $H^\pm$ and the DM candidate $H^0$, for which $\Delta M$ is below 1 GeV, the probable decay channels are mediated by off-shell $W^\pm$. As illustrated in the Fig.~\ref{fig:w-decays}, in the hadronic channel, $H^\pm \rightarrow H^0W^{\pm*}\rightarrow H^0\pi^\pm$, the off-shell $W^\pm$ hadronizes to a charged pion. Similarly, in the leptonic channel, $H^\pm \rightarrow H^0W^{\pm*}\rightarrow H^0\ell^\pm\nu_\ell$, the virtual $W^\pm$ decays into a charged lepton and the corresponding neutrino. However, just above the pion mass ($m_{\pi^\pm}\approx 140$ MeV), the hadronic decay channel is more dominant. As the final state contains a hadronic bound state, this decay cannot be described purely at the quark level using perturbative methods. Instead, it is described by the non-perturbative  $W-\pi$ mixing term in the lagrangian
 \begin{equation}
     \mathcal{L}_{W\pi}=\frac{gf_\pi}{2\sqrt{2}}W^+_\mu
\partial^\mu\pi^- +h.c.\,,
\end{equation}
where, $f_\pi=130$ MeV corresponds to the pion decay constant. In the limit $\Delta M /M_{H^\pm} \ll 1$, the decay width formula for the channel $H^\pm\rightarrow H^0\pi^\pm$ is given by~\cite{Belyaev:2016lok}
\begin{equation}
   \Gamma(H^\pm\rightarrow H^0\pi^\pm)=\frac{g^{4}\, f_{\pi}^{2}\, (\Delta M)^2}
{64~\pi~ M_W^{4}}
\sqrt{(\Delta M)^2 - m_{\pi}^{2}}\,.
\end{equation}
In the above, $g$ and $M_W$ are the $SU(2)_L$ gauge coupling and $W$ boson mass, respectively. Similarly, the three-body decay width for the leptonic channel is given by~\cite{Tsai:2019eqi}:

\begin{align}
\Gamma(H^\pm \to H^0 \ell \bar{\nu_\ell}) &= \frac{1}{120\pi^3}\frac{g^4}{M_W^4} (\Delta M)^5 
\sum_{jk} \left[|c_V^{jk}|^2 + |c_A^{jk}|^2\right] 
\times \theta(\Delta M^\pm - m_j - m_k)\,,
\end{align}
where, 
\begin{equation}
    c_V^{jk} = -c_A^{jk}=\frac{1}{2\sqrt{2}}\delta^{jk}~.
\end{equation}
In the above, $j,k$ runs over all flavours of SM leptons and the step function $\theta$ arises from the four momentum conservation. We can see from above that the decay width increases with $(\Delta M)$. This indicates a strong suppression in the decay width in the compressed regime. This results a long lifetime and displaced vertex signature of $H^\pm$. The left panel of Fig.~\ref{decay_width1} clearly explains the variation of decay width $\Gamma$ with mass splitting $\Delta M$ and the right panel illustrates the corresponding life-time $\tau$~(in $ns$) and decay length $c\tau$~(in $cm$) by the blue (thick) and red (dashed)  lines, respectively. The vertical dot-dashed lines indicate the kinematic thresholds corresponding to muon, pion and tau masses. The horizontal green dotted and purple dot-dashed lines denote characteristic dimension of CLIC tracking system, namely the outer tracker radius ($\sim 1.5$ m) and the radius of the inner layer vertex detector ($\sim 3$ cm). As evident from these figures, the lifetime of $H^\pm$ is strongly controlled by the mass splitting $\Delta M$. For a range of small $\Delta M \sim (140, 300)$ MeV, the corresponding decay length ($c\tau$) falls within the CLIC tracking volume, allowing the charged scalar to travel a measurable distance before decaying. This provides a natural parameter space for exploring long-lived charged-particle signatures. It is useful to note that the displaced vertex signal depends on $\Delta M$, unlike the dark sector, where both $\Delta M$ and $\delta m$ play a role. 


\section{Collider Phenomenology of the Long-Lived $H^\pm$}\label{collider_analysis}

The long-lived nature of $H^\pm$ provides a promising opportunity to probe the charged scalar through displaced vertex and disappearing track signatures at the colliders. In this section, we study the phenomenology of this long-lived $H^\pm$ at a future $e^+e^-$ collider with $\sqrt s=$ 3 TeV, motivated by its significantly larger production cross section compared to the LHC for all benchmark points considered here (see Table~\ref{tab:xsec}). The cleaner collider environment further makes $e^+e^-$ colliders suitable for probing such signatures of $H^\pm$.

\begin{figure}[!t]
 \centering 
\includegraphics[width=.5\textwidth]{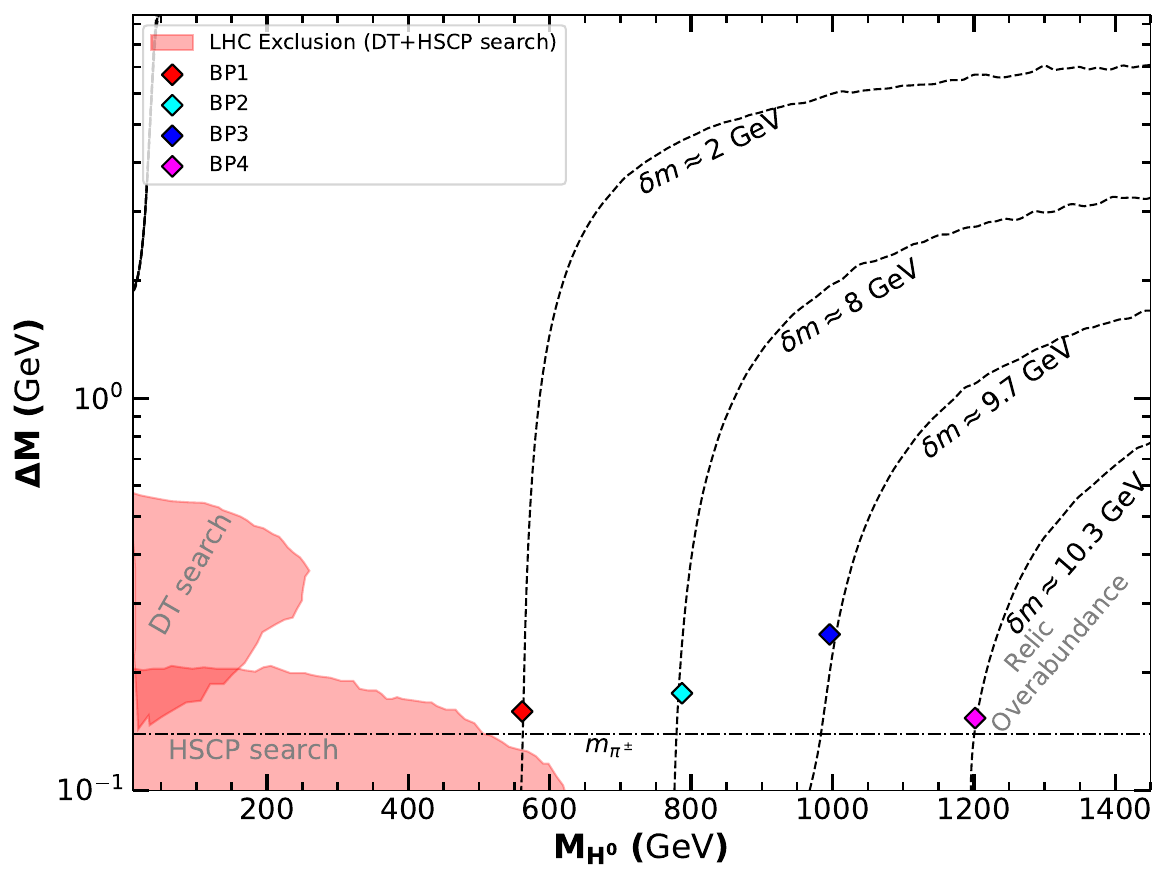}
 \caption{\label{smodels} LHC exclusion region (red) for the Heavy Stable Charged Particle (HSCP) and disappearing track (DT) searches. }
\end{figure}

For the analysis, we use the publicly available \textsc{Universal FeynRules Output} (UFO)~\cite{Darme:2023jdn} files for IDM\cite{Goudelis:2013uca} and generated the events at $e^+e^- \rightarrow H^+H^-$ followed by its hadronic decay at \textsc{MG5\_aMC\@NLO} (v3.5.5)~\cite{Alwall:2014hca}. We analyze the IDM parameter space by imposing dark matter analysis and current LHC limits. Based on these combined constraints, we identify four benchmark points with masses  ranging from around 550 GeV to 1.2 TeV. To evaluate the current LHC bounds, we use \textsc{SModelS\_v3}~\cite{Altakach:2024jwk} to recast the model predictions, which decomposes the signal into simplified model topologies. It excludes the parameter points depending on the value of $r$, with $r\equiv \sigma_{\rm theory}/\sigma_{\rm exp~ limit }.$ The value $r>1$ corresponds to exclusion of the corresponding parameter point. We present resulting exclusion bounds on the parameter spaces of IDM in the $M_{H^+}$ vs $\Delta M$ plane in Fig.~\ref{smodels}.  The red colored shaded region represents the excluded region on the plane derived from the different LHC searches, including CMS-EXO-13-006~\cite{CMS:2015lsu}, ATLAS-SUSY-2016-32, ATLAS-SUSY-2016-06~\cite{ATLAS:2017oal} and ATLAS-SUSY-2016-07~\cite{Bradmiller-Feld:2016yvd}. The first two analyses search for heavy stable charged particles (HSCP) while the third analysis searches for disappearing tracks arising from charged states and the fourth analysis searches for hadronic jets+MET ($\slashed{E}_T$). MET is identified to be the missing transverse momentum of the produced visible particles. 
\begin{equation}
   \slashed{E}_T=\sqrt{\sum_i{p_{x_i}}^2+ \sum_i{p_{y_i}}^2}\,,
\end{equation}
where $i$ runs over the visible particles produced in the event. 
The horizontal black  dot-dashed line in Fig.~\ref{smodels} denotes the pion mass. Below this line, the two body decay $H^\pm\rightarrow H^0\pi^\pm$ becomes kinematically suppressed. Here, the only the leptonic channels showing in Fig.~\ref{fig:w-decays}(b) becomes dominant and leads to even more long-lived charged states. However, in this study, we do not intend to explore the $m_{\pi^\pm}$ threshold.

In Fig.~\ref{smodels}, we also overlay our benchmark points on top of the select few $\delta m$ contours, which are safe from the current LHC exclusion bounds. The four benchmark points chosen for the analysis are detailed as in the Table~\ref{tab:BP1_lambdas} along with different observables and the quartic couplings. In Table~\ref{tab:xsec}, we tabulate the cross sections corresponding to the selected BPs at different colliders: at the CLIC , LHC13  and HL-LHC at the center of mass energy $\sqrt s$=3, 13 and 14 TeV respectively. .
\begin{table}[htbp]
\centering
\begin{tabular}{cccccc}
\hline
BPs&$M_{H^\pm}$ & $M_{H^0}$ & $M_{A^0}$  
&  $\Gamma(H^+\rightarrow H^0\pi^+)$ (GeV)& $c\tau~$ (cm) \\
\hline
BP1&561.047 & 560.888	&563.007  & $6.87\times10^{-16}$ &28.7\\
BP2& 788.874& 788.697 & 796.944  &$1.22\times10^{-15}$ & 16.1 \\
BP3&995.874 & 995.624  & 1005.544  & $4.67\times10^{-15}$&4.22 \\
BP4&1201.701 & 1201.548 & 1211.960  
&$5.21\times10^{-16}$ &37.8 \\ \hline


\hline
\end{tabular}
\label{tab:BP1}
\end{table}

\begin{table}[htbp]
\centering
\begin{tabular}{cccc}
\hline
BPs&  $\lambda_2$ & $\lambda_L$& $\sigma_{\text{SI}}^{p}(\rm{pb})$  \\
\hline
BP1 & 1.097 & 0.0090 & $9.03\times10^{-12}$\\
BP2 & 0.908 & 0.0098 &$5.94\times10^{-12}$\\
BP3 & 1.088 &  0.0039&
$5.46\times10^{-13}$\\
BP4 & 1.163 &0.01&
$2.65\times10^{-12}$\\ \hline


\hline
\end{tabular}
\caption{The benchmark points (BP1-BP4) of our model that satisfy all the existing bounds, along with the model parameters and different observables. The branching ratio of the decay channel ($H^\pm \rightarrow H^0\pi^\pm$) is $\sim 100\%$ for the chosen BPs. }
\label{tab:BP1_lambdas}
\end{table}



\begin{figure}[h]
 \centering 
\includegraphics[width=.49\textwidth]{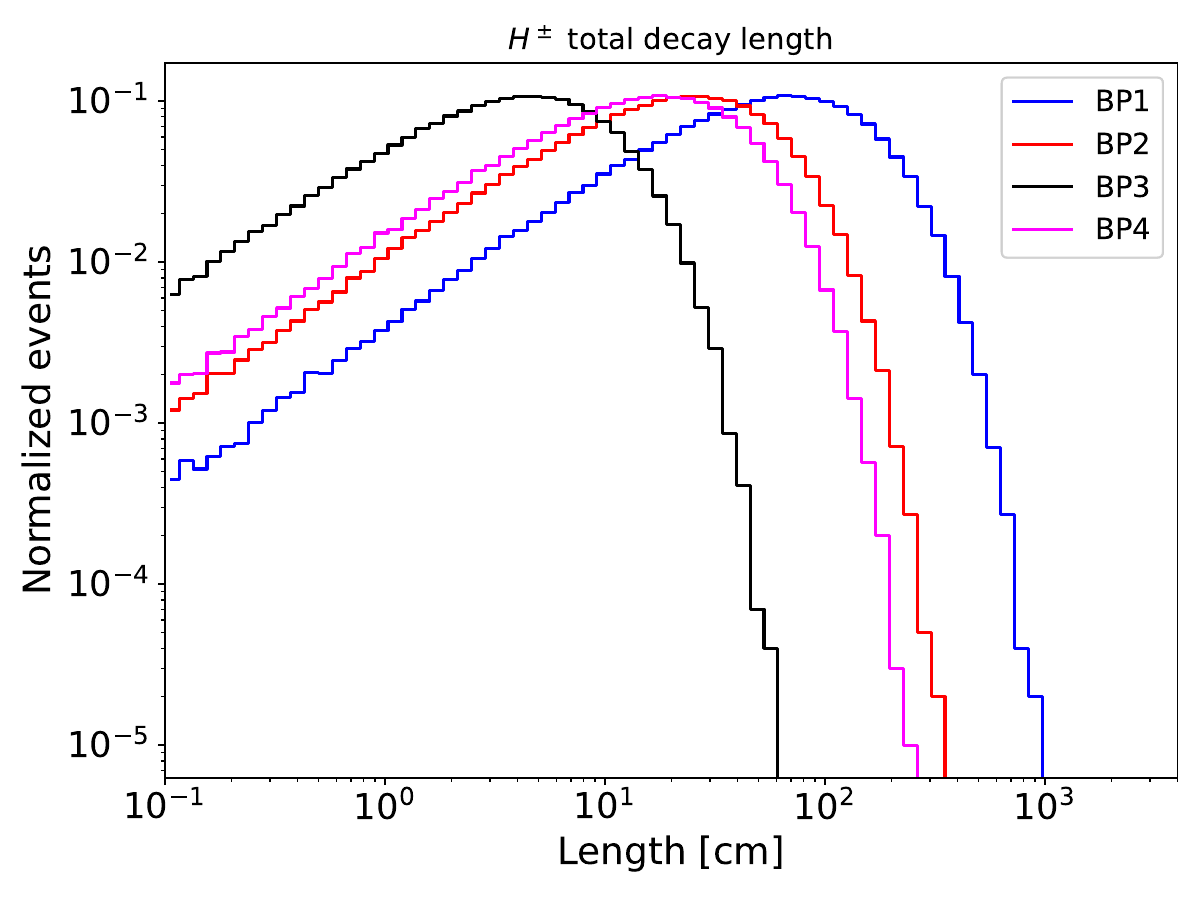}
\includegraphics[width=.49\textwidth]{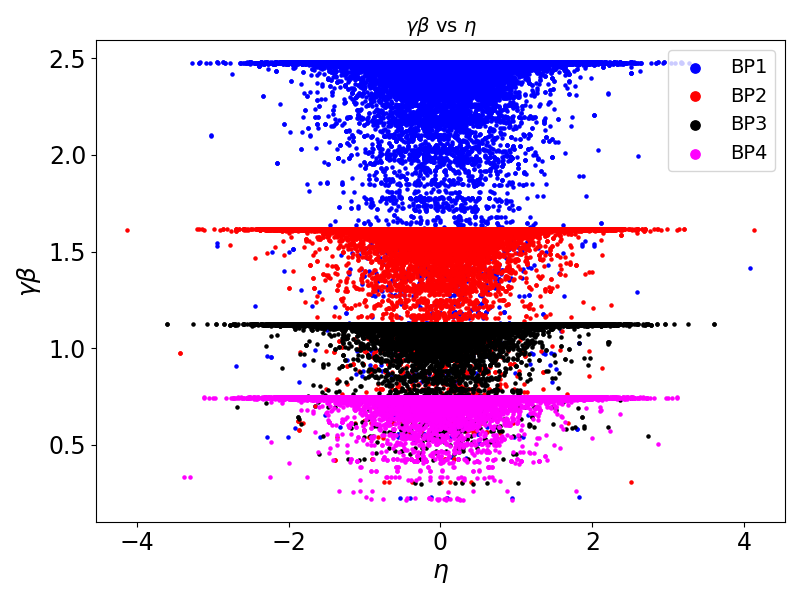}
\caption{\label{gammabeta} \textit{Left:}  The normalized distribution of the total decay length $L$ for the chosen benchmark points. The total decay length is given by $L=\langle L \rangle\rm{ln}(1-u)$, where $u$ is any random number in the range $[0,1]$ . \textit{Right} plot shows the distribution of the events in the $\eta-\gamma\beta$ plane.  }
\end{figure}

\begin{table}[h!]
\centering
\begin{tabular}{c c c c}
\hline\hline
\textbf{BPs} & \boldmath$e^{+}e^{-}$ & \textbf{LHC13} & \textbf{LHC14} \\
             & \textbf{(3 TeV)}      & \textbf{(13 TeV)} & \textbf{(14 TeV)} \\
\hline
BP1 & 2.61  & 0.185 & 0.228 \\
BP2 & 2.01    & 0.031    & 0.039    \\
BP3 & 1.37    & 0.007    & 0.010    \\
BP4 & 0.07 & 0.002 & 0.003 \\
\hline\hline
\end{tabular}
\caption{Pair production cross sections of $H^+H^-$ (in fb) for the benchmark points at future $e^{+}e^{-}$ collider and at the LHC with CM energies mentioned in the Table heading.}
\label{tab:xsec}
\end{table}
As a consequence of the small decay width, $H^\pm$ becomes long-lived with decay lengths ranging from a few centimeters to several tens of centimeters. This large $c\tau$ makes such scenarios favorable for the displaced vertex and disappearing track signatures at the colliders. To model the displaced decays of the charged Higgs $H^\pm$, we use  \textsc{Pythia8}(v8.3.12)~\cite{Bierlich:2022pfr} to simulate the mean boosted decay length  $L_{\rm boost}$, given by $L_{\rm boost}=\gamma \beta c \tau$. The actual decay position is then sampled from an exponential distribution $L=-L_{\rm boost}{\rm ln}(1-u)$, with $u\in [0,1]$ being uniformly distributed random number, thereby reproducing the expected probability density of decaying within the length $L$ 
\begin{equation}
    P(L)=\frac{1}{L_{\rm boost}}e^{-L/L_{\rm boost}}
\end{equation}
The total decay length is subsequently decomposed into transverse and longitudinal components using pseudo-rapidity $\eta$, yielding the radial decay position or the decay radius $R=L/\rm {cosh}\eta$ and longitudinal position $z=L{~\rm tanh}\eta$. Left panel of Fig.~\ref{gammabeta} shows the distribution of the total decay length $L$ (in cm) for the selected benchmark points. The right one shows the variation of the boosted factor $\gamma\beta$ with $\eta$.  

\begin{figure}[H]
 \centering 
\includegraphics[width=.49\textwidth]{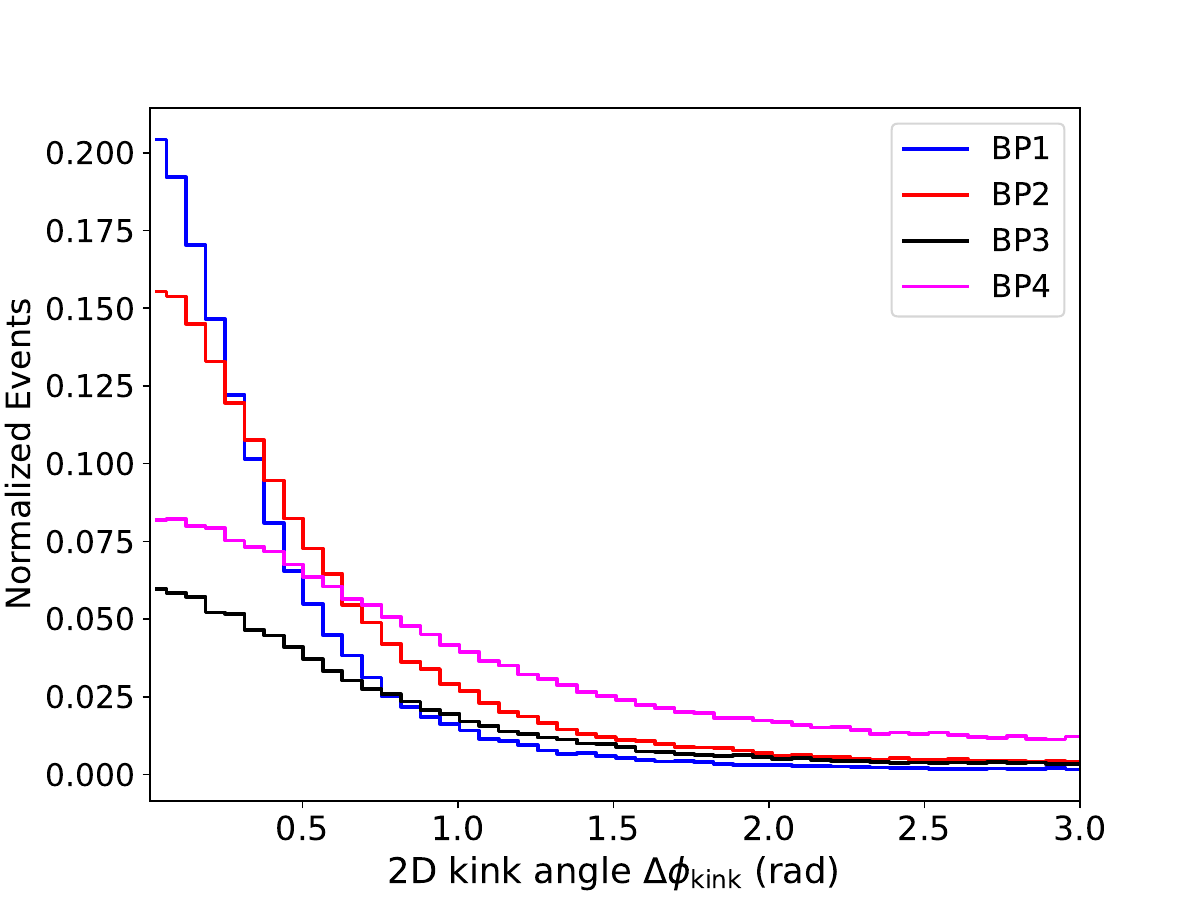}
\includegraphics[width=.49\textwidth]{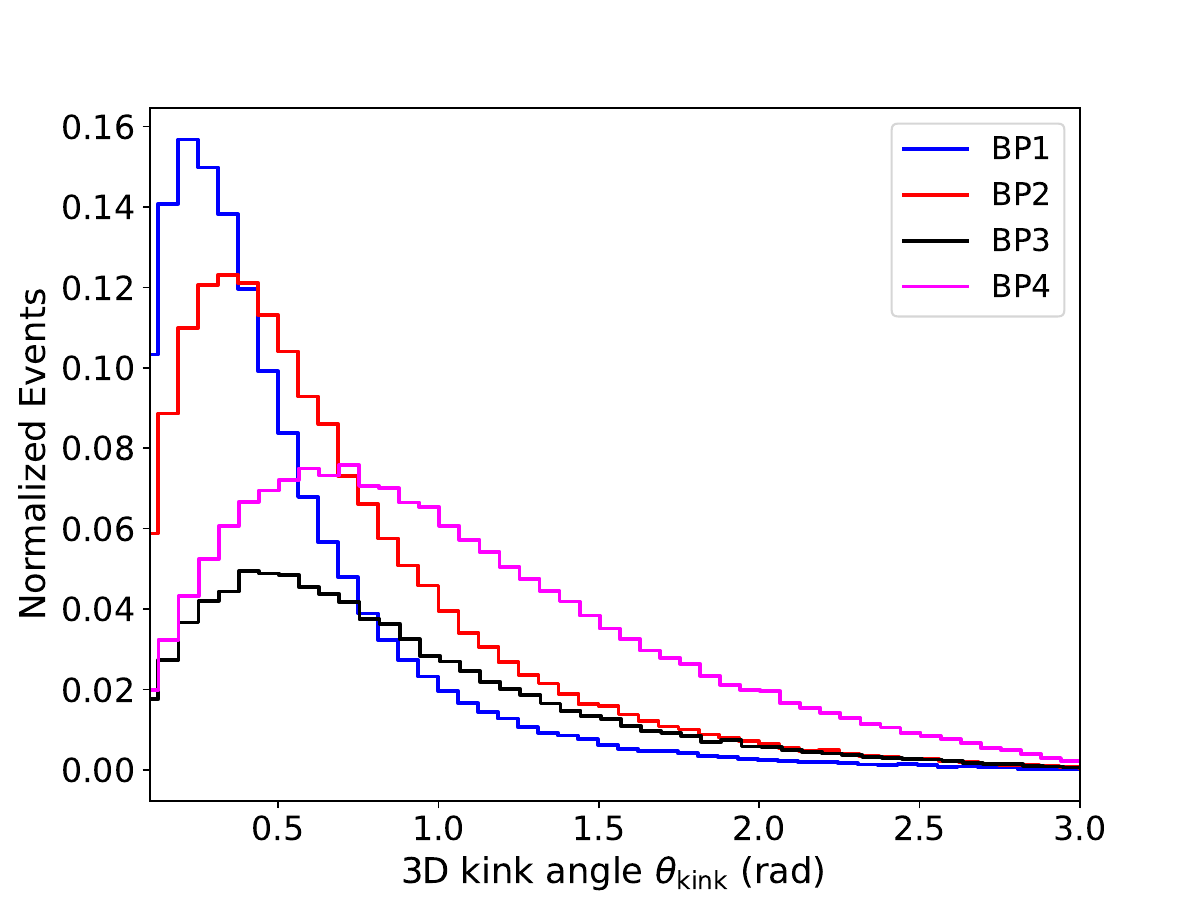}
\caption{\label{dedx} Two~(\textit{left}) and  three dimensional (\textit{right}) kink angle distribution between $H^+$ and $\pi^+$ with $3.5<R<100$ cm, $|\eta|<2.5$.}
\end{figure}

 To further characterize the signature of the long-lived $H^\pm$, we quantify the geometric deformation of the track arising from the displaced decay $H^\pm \rightarrow H^0+\pi^\pm$ through the three dimensional kink angle ($\theta_{\rm kink}$). This kink signature arises when a charged particle decays and produces a charged daughter particle, the track bends abruptly, which leads to a kink track. To observe the kink and to reconstruct the tracks separately, the kink must satisfy $\Delta \phi_{\rm kink}>0.1$, where, $\Delta \phi_{\rm kink}$ is the difference in azimuthal angles between mother and daughter tracks~\cite{Asai:2011wy}. As this angle is defined on the transverse plane, i.e perpendicular to the beam axis, we will term it as  2D kink angle. On the other hand, we also define the 3D kink angle for each event as :
 \begin{equation}
    \theta_{\rm kink}=  \rm{cos^{-1}}~\Big(\frac{\bar{p}_{H^\pm}. \bar{p}_{\pi}}{|\bar{p}_{H^\pm}||\bar{p}_{\pi}|}\Big)
 \end{equation}
Here, $\bar{p}_{H^\pm}$ and $\bar{p}_{\pi}$ denote the laboratory frame three momenta of $H^\pm$ before decay and the outgoing pion, respectively. This observable directly measures the angular deflection of the charged trajectory at the decay vertex inside the tracker. The first  and second plot of Fig.~\ref{dedx} shows the distribution of the 2D  and 3D kink angle $\theta_{\rm kink}$ for the selected benchmark points with $R\in[3.5~\rm cm , ~100~ cm]$ and $|\eta|<2.5$. 


\begin{figure}[h]
 \centering 
\includegraphics[width=.98\textwidth]{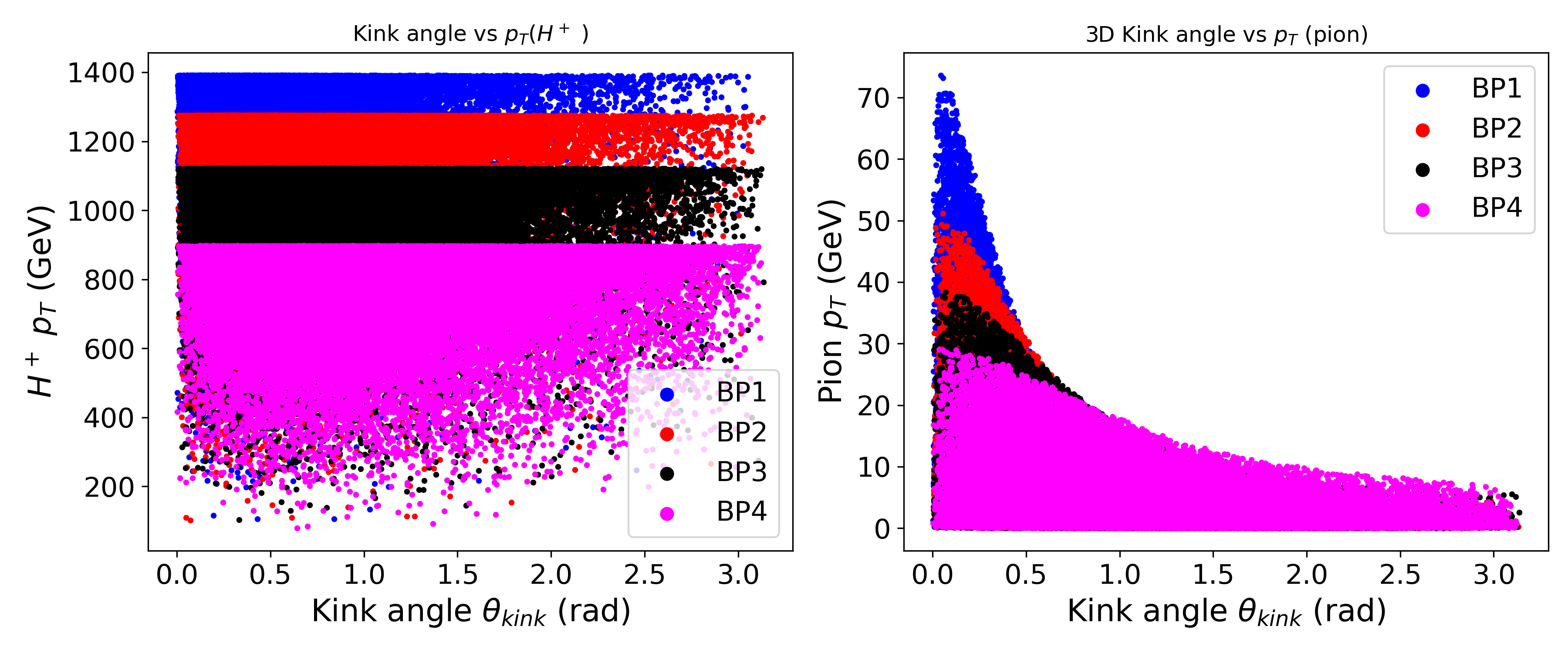}
\caption{\label{kink_cor} \textit{Left:} kink angle versus parent $H^+$ transverse momentum. \textit{Right:}  kink angle versus pion transverse momentum.}
\end{figure}
Using the above information, we plot the correlation between the kink angle and $p_T$ of the mother and daughter particles for our four benchmark points in Fig.\ref{kink_cor}. The left plot shows the variation of $\theta_{\rm kink}$ with the mother particle $p_T$ and right one shows with daughter $p_T$. It is evident from these plots that slow moving pions (small $p_T$) produce comparatively large kink angle.

\subsection*{Disappearing track of $H^\pm$:}
 As the charged Higgs $H^\pm$ travels, it leaves tracks within the tracking system, but due to their large mass, they deposit negligible energy. The track terminates inside the tracker when $H^\pm$ decays to soft visible states, giving a "kinked" and "disappearing trajectory". Such signal is rare in the SM and the main backgrounds for the disappearing track like signature come from the detector fakes i.e the backgrounds are entirely coming from the instrument. They emerge primarily from the catastrophic charged particle interaction with the detector material and from the random combinations of uncorrelated hits \cite{ATLAS:2017oal,ATLAS:2022rme}. However, in our study at $e^+e^-$ collider, we are not estimating these backgrounds, as they are highly detector specific, but assume some range of background events to plague the signal.
 
A reconstructed disappearing track requires the charged particle to leave sufficient hits in the tracking system, decay inside the tracker volume, and produce only soft visible decay products. To identify this topology, we apply a set of kinematic and geometric selection cuts. We choose the disappearing track candidate  required to decay within the tracker volume ($3.5 ~\rm cm<R < 100 ~\rm cm $). Further, we choose the 2D kink angle $\Delta\phi_{\rm kink}$ in the range $[0.1,~\pi/2]$. kink angle $\theta_{\rm kink}$ in between [0.1 rad, 2 rad]. 
\begin{figure}[h]
 \centering 
 \includegraphics[width=.49\textwidth]{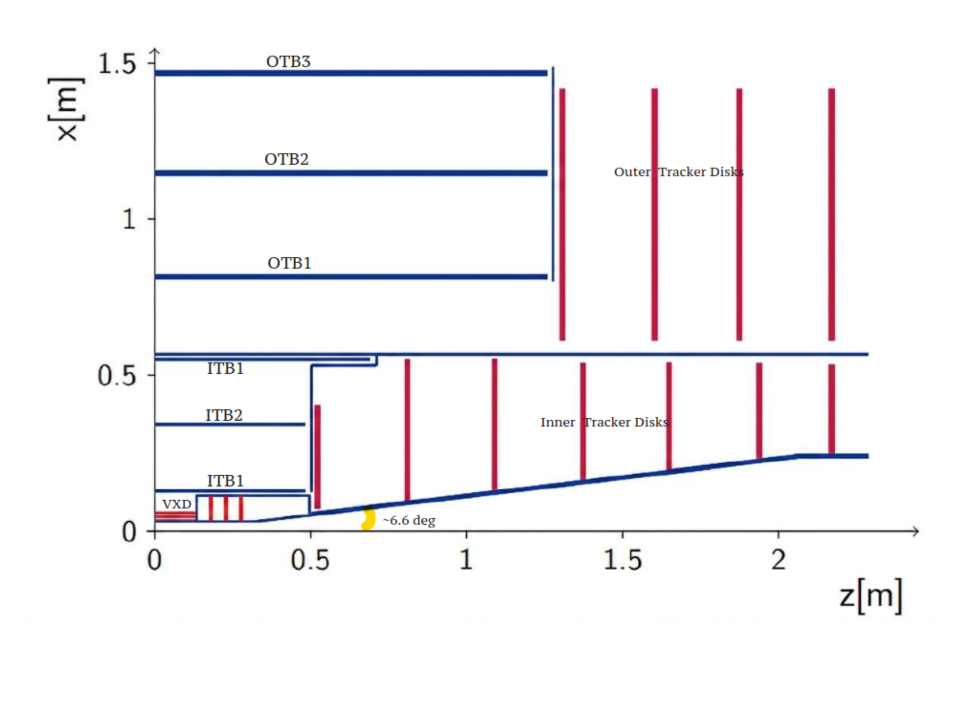}
\includegraphics[width=.49\textwidth]{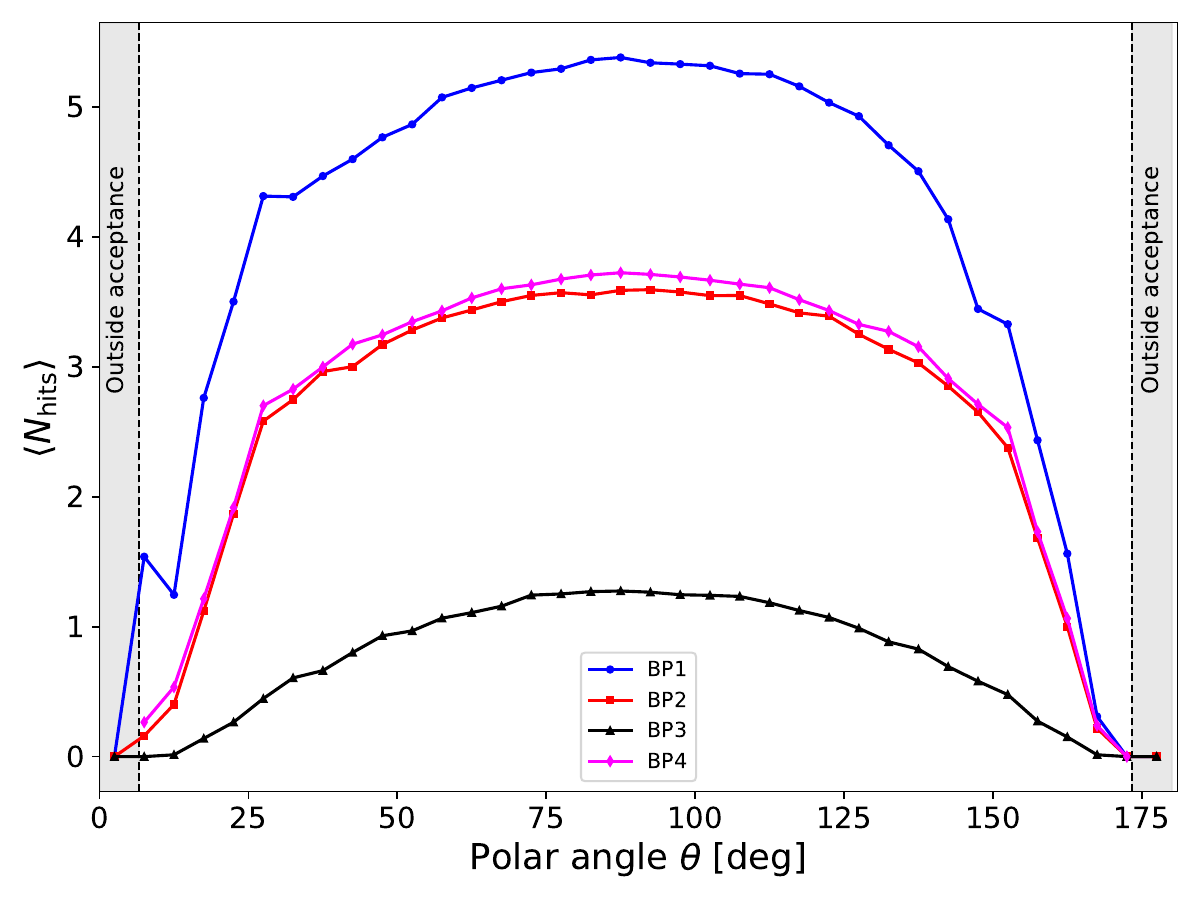}
\caption{\label{nhits}\textit{Left:} XZ view of the tracker of CLIC. Here, OTB, ITB and VXD stand for Outer Tracker Barrel, Inner Tracker Barrel and Vertex Detector, respectively. \textit{right:} Average number of tracking hits as a function of polar angle $\theta$ for the selected BPs.}
\end{figure}
To estimate the observability of the disappearing- track signature at the CLIC, we further evaluate the number of tracking hits associated with the long-lived charged scalar using the geometrical layout of the CLIC tracking system~\cite{CLICdp:2017vju}. The detector consists of 3 vertex detector (VXD) barrel layers, 6 silicon tracker barrel layers, 7 inner tracking tracking disks (ITD) and 4 outer tracking disks (OTD). The multiplicity is determined by checking whether charged scalar reaches a given detector layer before decaying. Fig.~\ref{nhits}  shows the average number of tracking hits ($\langle  N_{\rm hits}\rangle$ ), as a function of the polar angle $\theta$ for the four benchmark points. This hit multiplicity includes the contribution from VXD, tracker barrel layers and tracking disks. The shaded regions correspond to angular range outside the nominal CLIC tracking acceptance ($\theta<6.6^\circ$) and ($\theta>173.4^\circ$). This hit multiplicity shows a prominent dependency on the polar angle. In the central region, the charged Higgs particles travels a larger fraction of the barrel tracking system before decaying. This increases the number of registered hits in this region. Moving forward and backward regions, the particle intersects fewer barrel layers and hit pattern becomes dominated by the disk detectors. As a result, it decreases away from the central region. From the right plot of Fig.~\ref{nhits}, we see that maximum number of $\langle  N_{\rm hits}\rangle$ is achieved for BP1, while for BP3, it is the lowest. This follows from the fact that $p_T$ is maximum for BP1, and lowest for BP4 (see Fig.~\ref{kink_cor}). But $c\tau$ is larger for BP4 than BP3, which combines to produce the number of hits as inferred above.

 The details of the important cuts are presented in the Table \ref{tab:dt_cuts}. The first requirement demands the mother particle to decay within the tracker volume\footnote{As in the CLIC, the first vertex barrel layer is at a distance of 3.1 cm from the beam, therefore,  we choose the decay radius $R$ from around 3.5 cm.}. To identify the kink produced by the mother particle,  constraints are imposed on both $\phi_{\rm kink}$ and $\theta_{\rm kink}$. The lower limits suppress nearly collinear tracks, for which the decay can not be experimentally resolved, whereas the upper limits reject the events in which the daughter particle is moving backward. Finally, the disappearing charged track is required to leave at least four hits in different layers of the VXD and tracking system. This criterion guarantees a reliable reconstruction of the charged-particle trajectory and significantly reduces the fake-track backgrounds arising from detector noise or random hit combinations.






\begin{table}[H]
\centering
\begin{tabular}{|l|c|c|c|c|c|}
\hline
\textbf{Selection criteria} & \textbf{Cuts} & \multicolumn{4}{c|}{$\epsilon$} \\
\cline{3-6}
 &  & BP1 & BP2 & BP3 & BP4 \\
\hline

1. Decay radius & $3.5~\mathrm{cm} < R < 100~\mathrm{cm}$ 
& 0.74 & 0.84 & 0.43 & 0.84 \\

2. 2D Kink angle & $0.1 ~\rm rad < \Delta \phi_{\mathrm{kink}} < \frac{\pi}{2}$ 
& 0.48 & 0.58 & 0.30 & 0.56 \\

3. 3D Kink angle & $0.1~\mathrm{rad} < \theta_{\mathrm{kink}} < 2~\mathrm{rad}$ 
& 0.47 & 0.57 & 0.29 & 0.55 \\ 
 4. Num. of hits&$N_{\rm hits}\geq4$&0.38 & 0.37& 0.03 & 0.38\\ \hline
 Num. of signal events, $S$ & & 991 & 743 & 41 & 27 \\
  (at $\mathcal{L}=1~\rm ab^{-1}$) &&&&&\\

\hline
\end{tabular}
\caption{Selection criteria used to identify truth-level signal candidates with displaced decays of $H^\pm$ and kink-like topologies, together with the corresponding selection efficiencies
$\epsilon$ for the four benchmark points. Here,
$S=\sigma\times\epsilon\times\mathcal{L}$, where
$\sigma$ denotes the signal production cross section.}
\label{tab:dt_cuts}
\end{table}


The track efficiency is parameterized by 
\begin{equation}
\epsilon =
\frac{N_{\rm cuts}}
{N_{\mathrm{gen}}}\,,
\end{equation}
where $N_{\rm cuts}$ represents number of events after the cuts. Therefore, $\epsilon$ represents how many events survive after the cuts. In the Table~\ref{tab:dt_cuts}, we also see that a significant fraction of the signal events meet the requirements. Assuming an integrated luminosity of $1~\rm ab^{-1}$ at the 3 TeV CLIC, the expected number of signal events after all selection requirements are 991, 743, 41 and 27 for BP1, BP2, BP3 and BP4, respectively. This finding emphasizes the survival probability of the signal events and as a sensitive probe of the long-lived Higgs scenarios.

 \paragraph{Significance:} 
 \begin{figure}[h]
 \centering 
 \includegraphics[width=.45\textwidth]{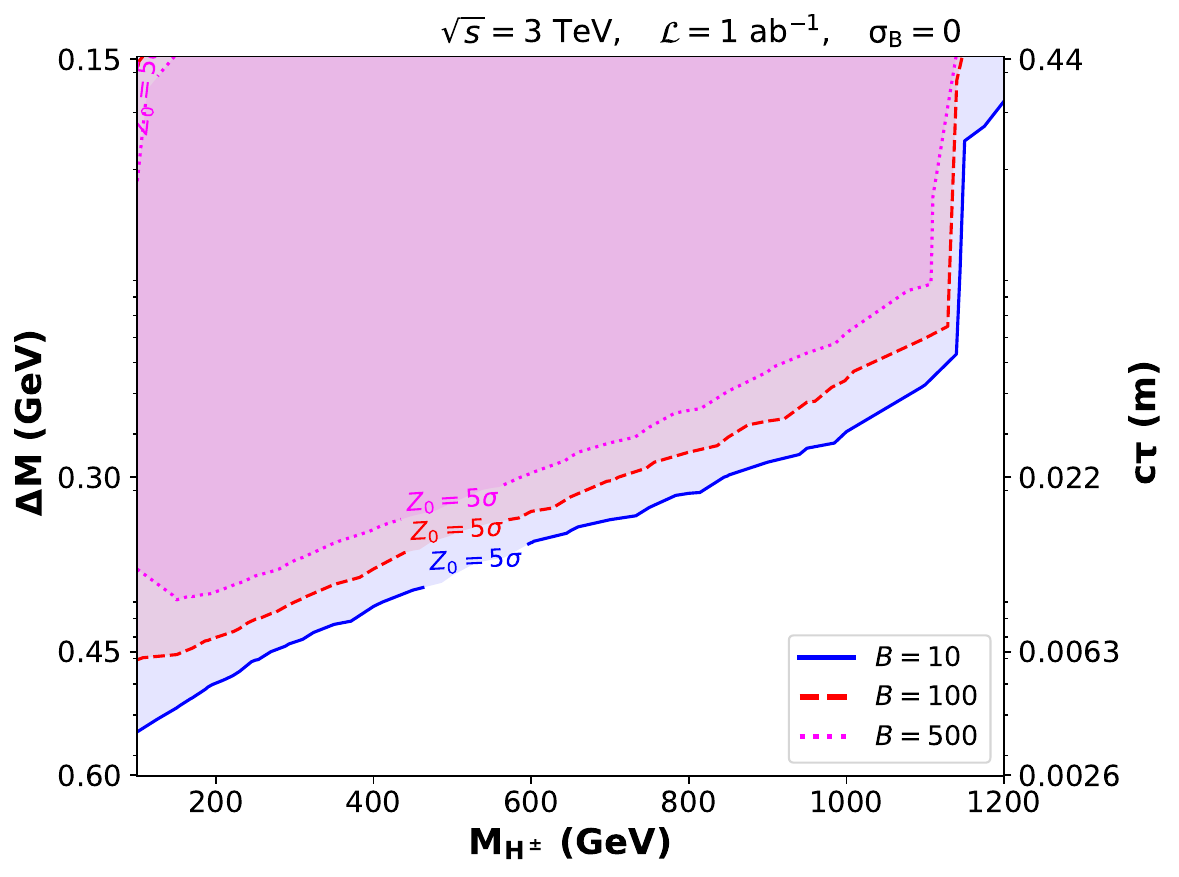}
 \includegraphics[width=.45\textwidth]{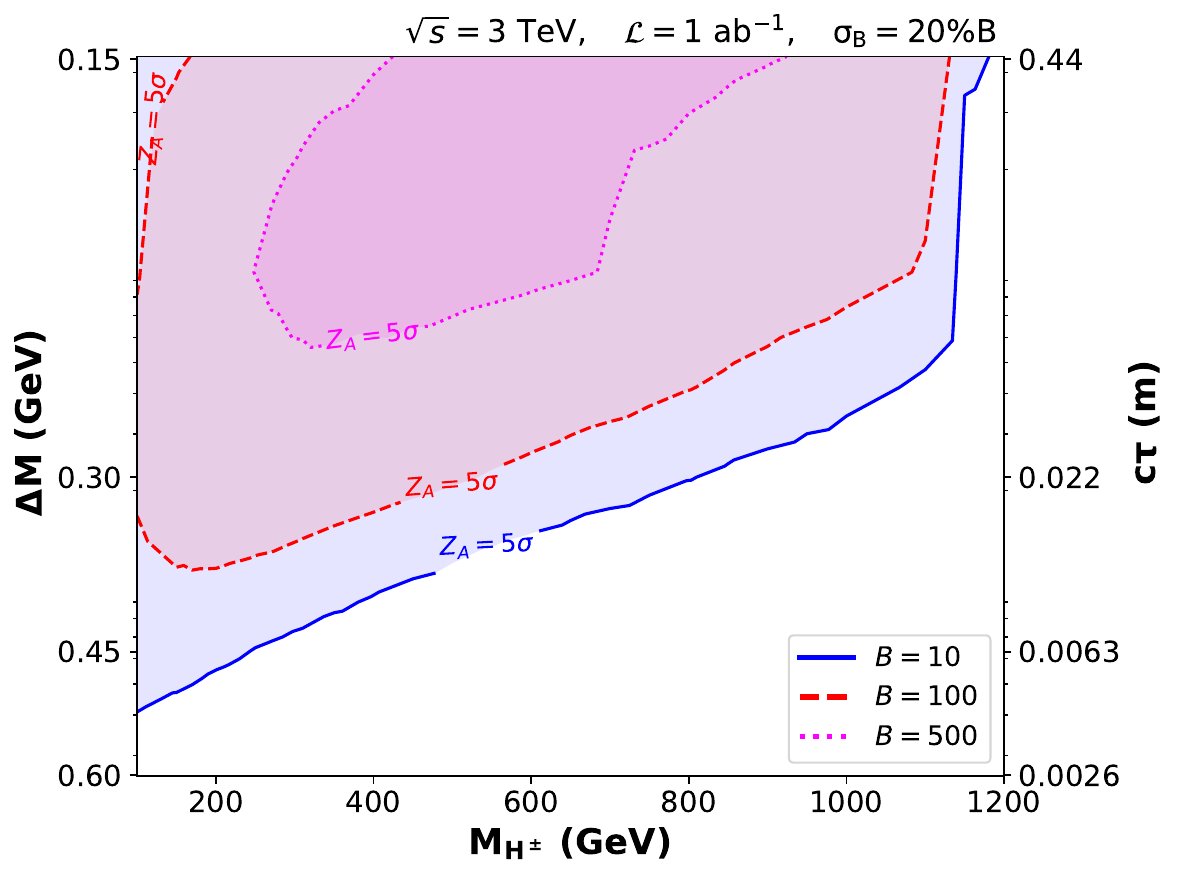}
\caption{\label{5sigma} The $5\sigma$ contours with systematic uncertainty $\sigma_B=0$ (\textit{left}) and with $\sigma_B=20\%$ in $B$ (\textit{right})  for representative background numbers $B=10,~100$ and 500. }
\end{figure}
To quantify the discovery potential of the signals, we evaluate statistical significance using two complementary measures: the Asimov significance including a systematic uncertainty on the background, $Z_A$, and the corresponding statistical-only significance, $Z_0$. Following the Asimov prescription of Ref.~\cite{Cowan:2010js}, the significance including a background uncertainty is given by:
\begin{equation}
\begin{split}
    Z_{A} = \sqrt{ 2 \left[ (S+B) \ln \left( \frac{(S+B)(B+\sigma_B^2)}{B^2+(S+B)\sigma_B^2} \right) - \frac{B^2}{\sigma_B^2} \ln \left( 1 + \frac{\sigma_B^2 S}{B(B+\sigma_B^2)} \right) \right]}\,,
\end{split}
\end{equation}
where, $S$ and $B$ denote the expected signal and background yields respectively,  and $\sigma_B$  is the uncertainty in the background yield, we use $\sigma_B=20\%$. In the limit where $\sigma_{B}$ tends to 0, the above expression reduces to
\begin{equation}
    Z_{0} = \lim_{\sigma_{B}\,\to\, 0} Z_A =\sqrt{2\left[(S+B)\ln{\left(1+\frac{S}{B}\right)}-S\right]}\,. 
\end{equation}
We begin by showing the $Z_0=5\sigma$ 
(left panel) and $Z_A=5\sigma$ (right panel)  contours in  Fig.~\ref{5sigma} as a function of mass of charged scalar ($M_{H^+}$), mass splitting ($\Delta\,M$) and decay length of charged mother particle for different values of background events. We choose three moderate to conservative values of background events viz. $10$, $100$ and $500$ with the uncertainty $\sigma_B=0$ and $\sigma_B=20\%$ in $B$, respectively. The corresponding $5\sigma$ contours are shown in the blue, red and magenta colors.  Within the SM, the process mimicking the signal topology is negligible and the majority of the signal contamination can arise from the detector effects, such as track reconstruction efficiencies, fake or misidentified tracks, and other instrumental effects. We therefore treat the number of background events as an effective parameter and investigate the robustness of the signal sensitivity against increasingly conservative background assumptions. An important feature of the Fig.~\ref{5sigma} is the non-monotonic dependence of the sensitivity on the mass splitting. The signal efficiency is strongly correlated with the decay length. For small $\Delta M$, the resulting long lifetime of $H^\pm$ causes its significant fraction to decay beyond the outer tracking boundary, whereas for large $\Delta M$, the shorter lifetime leads to decays before the minimum radial requirement $R_{\rm decay}\sim 3.5$ cm. Consequently, the sensitivity is maximized for an intermediate range of $\Delta M$. Obviously, on increasing the background from $B=10$ to 100 and to 500, reduces the $5\sigma$ sensitivity region, with the effect becoming more pronounced at larger $M_{H^\pm}$ due to the decreasing production rate. Thus, the projected sensitivity results from the interplay between the charged-scalar production rate, its life-time and the assumed background yield.

\begin{figure}[h]
 \centering 
 \includegraphics[width=.49\textwidth]{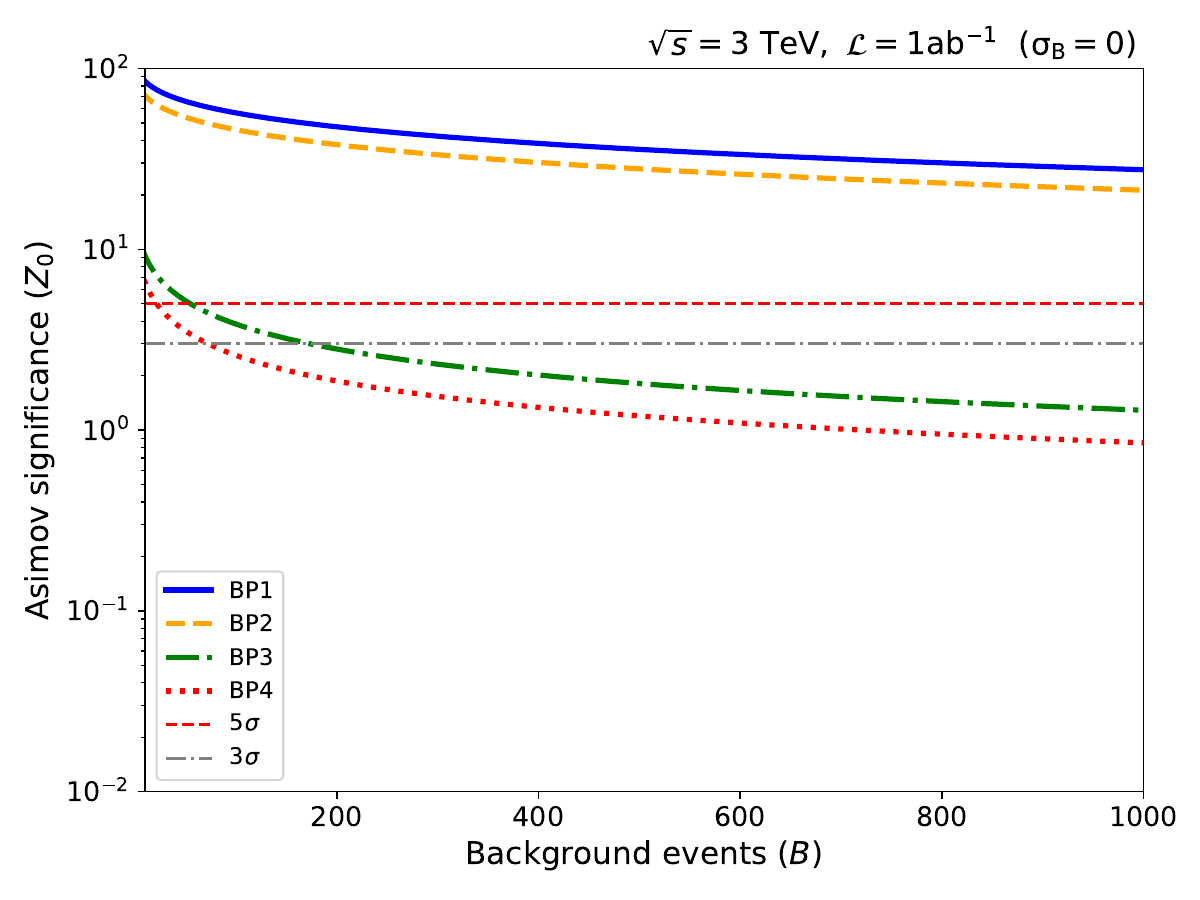}
\includegraphics[width=.49\textwidth]{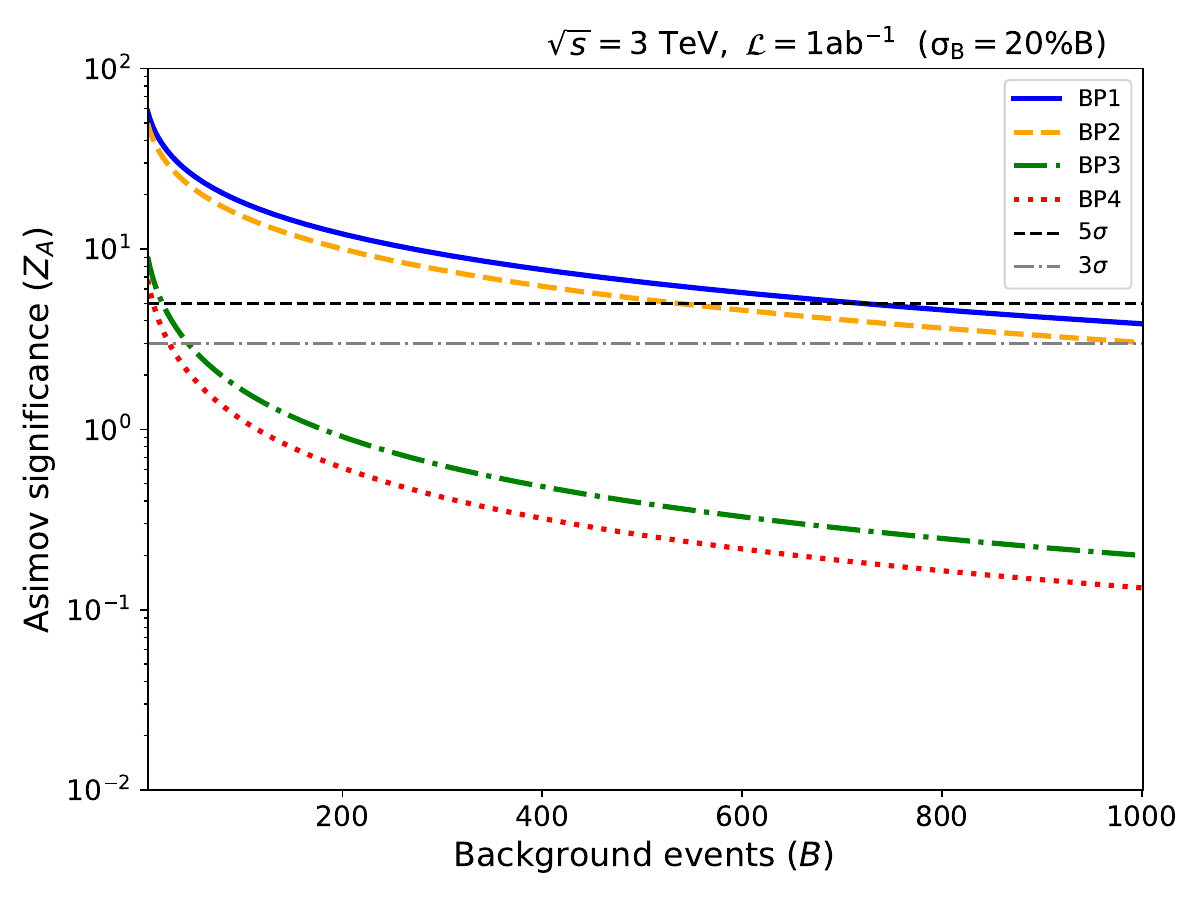}
\caption{\label{sigBP} \textit{Left:} Significance $Z_0$ as a function of expected number of backgrounds with systematic uncertainty $\sigma_B=0$. \textit{Right:} Significance($Z_A$)  with  $\sigma_B=$ 20\% . }
\end{figure}

Further, we estimate the significance of our model at the chosen representative benchmark points as a function of expected number of background events with and without systematic errors, as shown in Fig.~\ref{sigBP}. The left panel presents the statistical-only Asimov significance $Z_0$ assuming a zero systematic uncertainty $\sigma_B$ on the background, while the right panel shows the  significance $Z_A$ with $\sigma_B=20\%$. In both the plots, the benchmark points BP1, BP2, BP3 and BP4 are represented by the blue thick, orange dashed, green dot-dashed and red dotted lines, respectively. The horizontal red dashed and gray dot-dashed lines correspond to $5\sigma$ and $3\sigma$ line, respectively. The plots illustrate that the BP1 and BP2 have a sizeable significance with $ ~Z_0 \sim 20 \sigma ,~Z_A \sim 3\sigma$, even for a high background yield $B\sim 1000$. In contrast, BP3 and BP4 yield much smaller significances with $ Z_0 \sim \mathcal{O}(1)\sigma,~Z_A\sim \mathcal{O}(0.1)\sigma $. The comparison between $Z_0$ and $Z_A$ also clearly demonstrates the impact of assumed 20\% background systematic uncertainty. While the $Z_0$ remains relatively large for BP1 and BP2, the inclusion of systematic uncertainty substantially reduces the corresponding significance, particularly at larger background yields.

\section{Summary and Conclusion}\label{conclusion}
In this work, we explore the DM sector of the Inert Doublet model. We identified the parameter space where it satisfies the DM requirements and theoretical constraints. We mainly focus on the region with minimal mass splitting between the dark sector particles. We further recast these viable parameter region against the current LHC limits. On the basis of this, we picked four representative benchmark points that represent the main features of the allowed parameter space. \par
Next, we examine the collider prospects of these scenarios in the future $e^+e^-$ collider at $\sqrt{s}=3$ TeV. We study the disappearing track signature that results from a long-lived charged scalar. In case of a compressed mass spectra, the charged Higgs can have a long decay length. This leads to a track that ends when the mother particle decays to the soft visible products within the inner tracking volume. We use the event samples at the parton level and applied a series of kinetic and geometric selection criteria. These criteria include decay radius, number of hits, and kink angles. These help us to characterize the signal and get the signal efficiencies. Our results show that a noticeable fraction of signal events meet the selection requirements across the chosen benchmark points. This illustrates the potential probe of disappearing track searches to the IDM parameter space in the compressed regime. The clean environment of the lepton collider also help these signatures to observe by reducing the backgrounds. However, we stress this study in the very parton level and do not include any detector simulations or reconstruction effects. Specifically, we did not include the instrumental backgrounds that come from the fake detector hits. This needs a dedicated detector-level analysis. Therefore, the efficiencies we present should be seen as an optimistic estimate of the signal reach. \par
In summary, our study highlights the potential of disappearing track signature in probing long-lived charged particles belonging to the dark sector. We have explicitly demonstrated for IDM scenario at future lepton colliders with unpolarized beams, where the charged scalar is heavy and decays to the dark matter and soft visible partners, naturally abiding by the DM constraints. The promise of disentangling such  charged track signals at future lepton colliders also stem from larger signal cross-section of heavy charged particles appearing in electroweak extension, and less background contamination than that of LHC, indicating the need to build such a collider. The methodology adopted in this analysis can easily be applied for other extended dark sector scenarios with compressed spectra.

\section*{Acknowledgements}
SB acknowledges the Anusandhan Research Foundation (ANRF) grant CRG/2023/000580. HD would like to thank the Council of Scientific \& Industrial Research (CSIR), Govt. of India for the senior research fellowship.

\appendix
\bibliographystyle{JHEP}
\bibliography{refer}

\end{document}